# Morphotropic Phase Boundary (MPB) Induced Enhancement of Ferroelectric and Piezoelectric Properties in Li and Ta modified $K_{0.5}Na_{0.5}NbO_3$

Satyaranjan Sahoo[1], Dhiren K. Pradhan[2*], Shalini Kumari[3], Abhisikta Sahu[1], Koyal Suman Samantaray[4], Vikas N. Thakur[5,6], Anupam Mishra[7], M. M. Rahaman[8], Ashok Kumar[5,6], Reji Thomas[9,10], Philip D. Rack[2], and Dillip K. Pradhan[1*]

[1]Department of Physics and Astronomy, National Institute of Technology Rourkela, Rourkela, Odisha 769008, India

[2] Department of Electrical and Systems Engineering, University of Pennsylvania, Philadelphia, PA, 19104, USA.

[3]Department of Materials Science & Engineering, The Pennsylvania State University, University Park, Pennsylvania 16802, USA

[4]Department of Physics, Indian Institute of Technology Indore, Indore, 453552, India

[5]CSIR-National Physical Laboratory, Dr. K. S. Krishnan Marg, New Delhi 110012, India

[6]Academy of Scientific and Innovative Research (AcSIR), Ghaziabad 201002, India

[7]Department of Ceramic Engineering, National Institute of Technology Rourkela, Rourkela, Odisha 769008, India

[8]Department of Materials Science and Engineering, University of Rajshahi, Rajshahi 6205, Bangladesh

[9]Division of Research and Development, Lovely Professional University, Jalandhar-Delhi G.T. Road, Phagwara, Punjab 144411, India

[10]School of Chemical Engineering and Physical Sciences, Lovely Professional University, Jalandhar-Delhi G.T. Road, Phagwara, Punjab 144411, India

[*] Author to whom correspondence should be addressed: dillip.pradhan79@gmail.com, pdhiren@seas.upenn.edu

## Abstract

Lead-free $(K_{0.48}Na_{0.48}Li_{0.04})(Nb_{1-x}Ta_x)O_3$ (KNLNT-$x$) ceramics were synthesized and the effect of Li and Ta substitution on the phase transition behavior, microstructure, ferroelectric, dielectric and piezoelectric properties were systematically investigated. X-ray diffraction and Raman spectroscopic analysis reveals that the KNLNT-$x$ ceramic with $x < 0.10$ exhibits a single phase *Amm2* crystal structure. A coexistence of orthorhombic and tetragonal (O+T) dual phase (*Amm2*+ *P4mm*) was observed in the composition range of $0.10 \leq x \leq 0.20$. With increasing $x$ value the structure changes from the O+T coexistence region to a single phase tetragonal (*P4mm*) structure for $x > 0.20$. Electron microscopy shows a highly dense ceramic with inhomogeneous distribution of grains and decreasing grain size with increasing Ta concentration. Temperature-dependent dielectric properties exhibit two phase transitions: the orthorhombic to tetragonal ($T_{O-T}$) phase transition and tetragonal to cubic phase transition ($T_C$). With increasing Ta concentration, $T_{O-T}$ and $T_C$ systematically reduces, the transition peaks broaden, and $T_{O-T}$ shifts below room temperature for $x > 0.15$. The KNLNT-0.20 shows the highest dielectric constant ($\varepsilon_r$

= 556) and piezoelectric coefficient ($d_{33}$ = 159pC/N). The enhanced piezoelectric property of KNLNT-0.20 is attributed to the morphotropic phase boundary composition rather than shifting of the polymorphic phase boundary (PPB) temperature towards room temperature. Based on the RT XRD data, Raman spectra, and temperature-dependent dielectric properties, a phase diagram has also been constructed.

# I. INTRODUCTION

In some of the non-centrosymmetric dielectric materials, electrical and mechanical properties are coupled to realize piezoelectricity that allows the conversion of mechanical energy to electrical energy and vice versa [1,2]. Among such materials, some show pyroelectricity, a temperature-dependent polarization phenomenon. Ferroelectrics have non-linear dielectrics properties with stable and electrically switchable polarization, and is the subgroup of piezoelectric and pyroelectric materials [1]. Piezoelectric, pyroelectric, and ferroelectric materials are widely used in various electronic and optoelectronic devices such as actuators, transducers, sensors, ultrasonic medical imaging, mechanical energy harvesting, structural health monitoring, etc., due to their excellent physical properties [1,2]. Piezoelectric/ferroelectric materials have been widely used for device applications, since the discovery of ferroelectricity in $BaTiO_3$ ceramic oxides during World War II [2]. However, at present, much research have been carried out to further improve the functional ferroelectric properties. Presently, many devices manufactured by various electronic industries are dominated by lead-based ferroelectric materials such as $Pb(Zr_xTi_{1-x})O_3$ (PZT), $Pb_{1-x}La_x(Zr_yT_{1-y})_{1-x/4}O_3$ (PLZT), $(1-x)Pb(Mg_{1/3}Nb_{2/3}O_3)$-$xPbTiO_3$ (PMN-PT), which contains more than 60 *wt.*% of lead [1-6]. The toxicity of lead and its high vapor pressure during synthesis/processing are harmful to humans and the environment [3-5]. Legislation has been enacted by the European Union for the restricted use of the Pb-based ferroelectric system in electronic devices [3, 4, 6], thus, there is a need for lead-free ferroelectric systems. $BaTiO_3$ (BT), $(Na_{0.5}Bi_{0.5})TiO_3$ (NBT), $(K_{0.5}Bi_{0.5})TiO_3$ (KBT), $(K_{0.5}Na_{0.5})NbO_3$ (KNN) based systems have been developed by different research groups around the globe [2-10], however, these materials are yet to be used in various devices by the electronic industry.

Among the aforementioned lead-free ferroelectric systems, KNN-based ferroelectric materials have received more attention after the reported work of Saito *et al.* on Li, Ta, and Sb co-doped KNN textured system, which showed the piezoelectric coefficients as high as 416 pC/N [11]. KNN has a high Curie temperature ($T_C$ ~ 420 °C), and good ferroelectric properties with a high remnant polarization ($P_r$ = 33 $\mu C/cm^2$), which are appealing for device applications [8, 12]. However, KNN prepared under ordinary conditions shows relatively lower piezoelectric properties ($d_{33}$ ~ 80 pC/N) due to difficulty in obtaining high density [8, 13]. Thus, tremendous efforts have been made to enhance the physical properties by optimizing the synthesis condition and chemical composition modifications [14]. KNN with the morphotropic phase boundary

(MPB) having the composition of 50 mol. % K and 50 mol. % Na, separating two orthorhombic phases has been realized [15,16]. The structural phase transition behavior of the KNN system is similar to the BaTiO$_3$ ferroelectric system, where at room temperature KNN crystallizes in an orthorhombic crystal structure with an *Amm2* space group (SG). The ferroelectric transition (Curie temperature, T$_C$) is around 420 °C, and the material becomes paraelectric as the structure changes to the cubic with $Pm\bar{3}m$ space group [4, 6, 9]. With decreasing temperature, the cubic paraelectric unit cell undergoes a deformation to ferroelectric tetragonal structure with SG: *P4mm* at 420 °C. Further cooling below 200 °C (T$_{O-T}$), it transforms to orthorhombic structure with SG: *P4mm,* and finally, it transforms to *R3m* symmetry with rhombohedral crystal structure also having ferroelectricity below -150 °C (T$_{O-R}$) [4, 6, 9]. The phase boundary of rhombohedral to orthorhombic (T$_{O-R}$) and orthorhombic to tetragonal (T$_{O-T}$) phase transitions depends on both temperature and composition, hence known as a polymorphic phase boundary (PPB) [6].

As described above, the major problems associated with the KNN system is the synthesis of high-quality KNN system because (i) it needs special handling of the starting precursors (ii) its physical properties are very sensitive to the stoichiometry and (iii) the complex densification process associated results in low density [15, 16]. As a result, different authors have reported different values of piezoelectric and electro-mechanical coupling for the same composition. To enhance the density and suitable microstructure, various synthesis/optimization techniques such as (i) addition of a sintering aid, (ii) hot pressing, (iii) spark plasma sintering, (iv) microwave sintering, (v) controlled atmosphere sintering, (vi) chemical modification (doping, solid solutions) have been adopted widely [6, 9, 15]. Hence, effort has been made to develop a variant KNN-based system with improved piezoelectric and electro-mechanical coupling. It is also believed that the high piezoelectric and electromechanical coefficients are due to the induction of a morphotropic phase boundary (MPB) over a small composition range in the ferroelectric systems [13, 17]. Further, to improve the piezoelectric and other physical properties of KNN by inducing the MPB at RT, chemical modification approaches have been widely used [6].

Although the enhanced piezoelectric properties in KNN modified systems were previously attributed to the formation of MPB (like PZT system), a few reports also suggested the enhanced piezoelectric properties are due to the shifting of T$_{O-T}$ towards the room temperature [6, 18-20]. It should be mentioned here that, the PPB results in a strong temperature-dependent piezoelectric, ferroelectric, and dielectric properties, which is not suitable for many device applications [18]. A clear understanding of the MPB separating the orthorhombic and tetragonal phases in case of KNN-based systems has not been evolved to definitively assign the large piezoelectric and electromechanical coupling to the MPB effect [15]. Although many studies have been carried out, the piezoelectric properties of KNN are still inferior to PZT [21].

In order to increase the piezoelectric and other physical properties, different solid solutions of KNN with other ferroelectrics perovskites such as Bi$_{0.5}$K$_{0.5}$TiO$_3$, Bi$_{0.5}$Na$_{0.5}$TiO$_3$, Bi$_{0.5}$Li$_{0.5}$TiO$_3$, BiAlO$_3$, BiScO$_3$, BiFeO$_3$, BaTiO$_3$, SrTiO$_3$, SrZrO$_3$, CaTiO$_3$, LiNbO$_3$, LiSbO$_3$, LiTaO$_3$ have been synthesized [6,11-13,18,21-28]. There are various reports where dopants like Li$^+$, Ag$^+$ are substituted at the K/Na site and dopants like $Ta^{5+}$, $Sb^{5+}$, $Zr^{4+}$, $Hf^{4+}$ are substituted at the Nb site of KNN [4, 6]. In this manuscript, we have substituted $Li^+$ at the A-site and $Ta^{5+}$ at the B-site of KNN [6, 8, 15, 17, 21, 22]. Readers are referred to Supplementary Table 1 for a brief description of the various Li- or/and Ta-modified KNN ceramics [6, 8, 13-15, 17, 18, 21, 22, 29-37].

Guo *et al.* have reported that the MPB region of KNN due to the existence of orthorhombic and tetragonal phases plays an important role in the enhancement of the piezoelectric properties [13]. The substitution of Li (6 mol.%) at the A-site of KNN system can induce a MPB with orthorhombic and tetragonal phases, as a result, enhanced piezoelectric and electromechanical coupling is observed around the MPB region [15]. $Li^+$ ion substitution at the A-site of KNN suppresses the formation of the low-temperature rhombohedral phase [8]. Substitution of Li at the A-site of KNN is very effective in enhancing piezoelectric properties and also increased the $T_C$ as well as decrease the $T_{O-T}$. Thus, the interval between both the phase transitions i.e. the stability of the temperature range of tetragonal phase increases [21]. The substitutions of $Li^+$ (4 to 20 mol%) at the A-site of KNN showed improved piezoelectric properties. However, above 8 mol. % Li substitution in KNN leads to the beginning of the formation of pyrochlore phases (i.e. $K_3Li_2Nb_5O_{15}$) [8]. Similarly, Ta-substitution also showed enhanced piezoelectric properties as compared to pure KNN systems [15]. Higher density and higher piezoelectric properties have been obtained by Li and Ta substitution using the ordinary synthesis techniques [15]. Simultaneous substitution of Ta and Sb is also responsible for the further enhancement of the dielectric and piezoelectric properties. The high electro-negativity of Ta compared to Nb may lead to more covalent rather than ionic bonding and results in enhanced physical properties. Li substitution increases $T_C$ and decreases $T_{O-T}$, however, Ta substitution decreases both $T_C$ and $T_{O-T}$ [15]. As the orthorhombic phase possesses large number of possible domain states, it may lead to higher strain-electric field hysteresis [15]. Hence we have restricted Li substitution to 4 mol. % to maintain the orthorhombic phase without Ta substitution.

Various reports are available on the enhancement of the physical properties of the KNN system due to the existence of MPB, which are attributed to the existence of different crystallographic phases [20]. It has also been reported that the improvement in the physical properties in KNN-based systems is due to the shifting of the polymorphic phase boundary (PPB i.e., $T_{O-T}$) close to RT rather than the existence of a MPB [38]. Due to the shifting of PPB towards RT, the piezoelectric and ferroelectric properties are strongly dependent on temperature; hence it would not be suitable for device applications. Based on the literature, two different mechanisms for the enhancement of the physical properties have been reported. i.e. (1) induction of MPB due to chemical modification and (2) the shifting of the orthorhombic-tetragonal polymorphic phase boundary towards room temperature. Therefore phase coexistence of O+T phase in case of KNN system is the primary criteria for both PPB and MPB, however, the origin of the PPB and MPB are quite different. Although both of these mechanisms have been reported to be the enhancement of physical properties, it is necessary to understand the difference between the MPB and PPB in the KNN system to corroborate the mechanism of the enhancement of the physical properties. The difficulty in understanding the structural and microscopic origin (despite of years of studies) of the enhancement of piezoelectric and ferroelectric properties in case of KNN system has motivated us to systematically study the effect of Li and Ta substitution on physical properties and thoroughly correlate this with crystal structure.

## II. EXPERIMENTAL DETAILS
### A. Synthesis Procedure

The lead-free ferroelectric ceramics of $(K_{0.48}Na_{0.48}Li_{0.04})(Nb_{1-x}Ta_xO_3)$ (KNLNT-$x$, where $x$ = 0.00, 0.05, 0.10, 0.15, 0.20, 0.25, 0.30, 0.40) were prepared by a conventional solid state reaction technique. High purity metal carbonates of $K_2CO_3$ (99% Alfa Aesar), $Na_2CO_3$ (99.5% Sigma Aldrich), $Li_2CO_3$ (99%) and metal oxides of $Nb_2O_5$ (99.99% Sigma Aldrich) and $Ta_2O_5$ (99.85% Alfa Aesar) were used as the precursors. Prior to mixing, $K_2CO_3$ and $Na_2CO_3$ were dried at 220 °C for 4h to remove any moisture intake owing to their hygroscopic characteristics. The precursors were then weighed according to the desired stoichiometry of the selected compositions and mixed for 3h using an agate mortar in both dry and acetone medium. A calcination process of the ground powders was then carried out at 875 °C for 6h in air medium. The calcined powders were again ground, mixed thoroughly with 3 wt. % polyvinyl alcohol (PVA) solution and then pressed into the disks of 10 mm diameter and 1.3 mm thickness. The pellets were finally sintered at optimum sintering temperature between 1100 °C – 1180 °C for 3h in air medium for densification.

### B. Characterization Techniques

X-ray diffraction (XRD) measurement was performed using an X-ray diffractometer, (Bruker D2-Phaser) with Cu $K_\alpha$ radiation ($\alpha$ = 1.5406 A°). The room temperature (RT) XRD data of the powdered samples were collected over a wide range of Bragg's angle (2θ) 20° to 100° at a scanning rate of 2°/min with 0.02° step size. Raman spectroscopic measurement was carried out at room temperature using a micro Raman spectrometer (Invia, Renishaw, UK). The surface morphology and the process of grain growth in ceramic samples were studied using Scanning Electron Microscopy (SEM-JEOL-JSM 6480 LV). The temperature-dependent dielectric properties of the samples were measured using an impedance analyzer (HIOKI IM3570) over the frequency range 100 Hz to 1 MHz. The RT ferroelectric hysteresis (P-E) loop of the ceramics was performed using a Ferroelectric loop tester (Radiant Ferroelectric Tester) at 5 Hz. The piezoelectric coefficient $d_{33}$ was measured for the poled samples at room temperature with the help of $d_{33}$ piezometer (APC International Ltd). Prior to the electrical measurement of the sintered samples, both sides of the pellets were polished, painted with silver paste and then dried at 200 °C for 2h to remove any moisture. The poling of the samples was then carried out at room temperature in a silicon bath by the application of a DC electric field of 2.5 kV/mm for 12h.

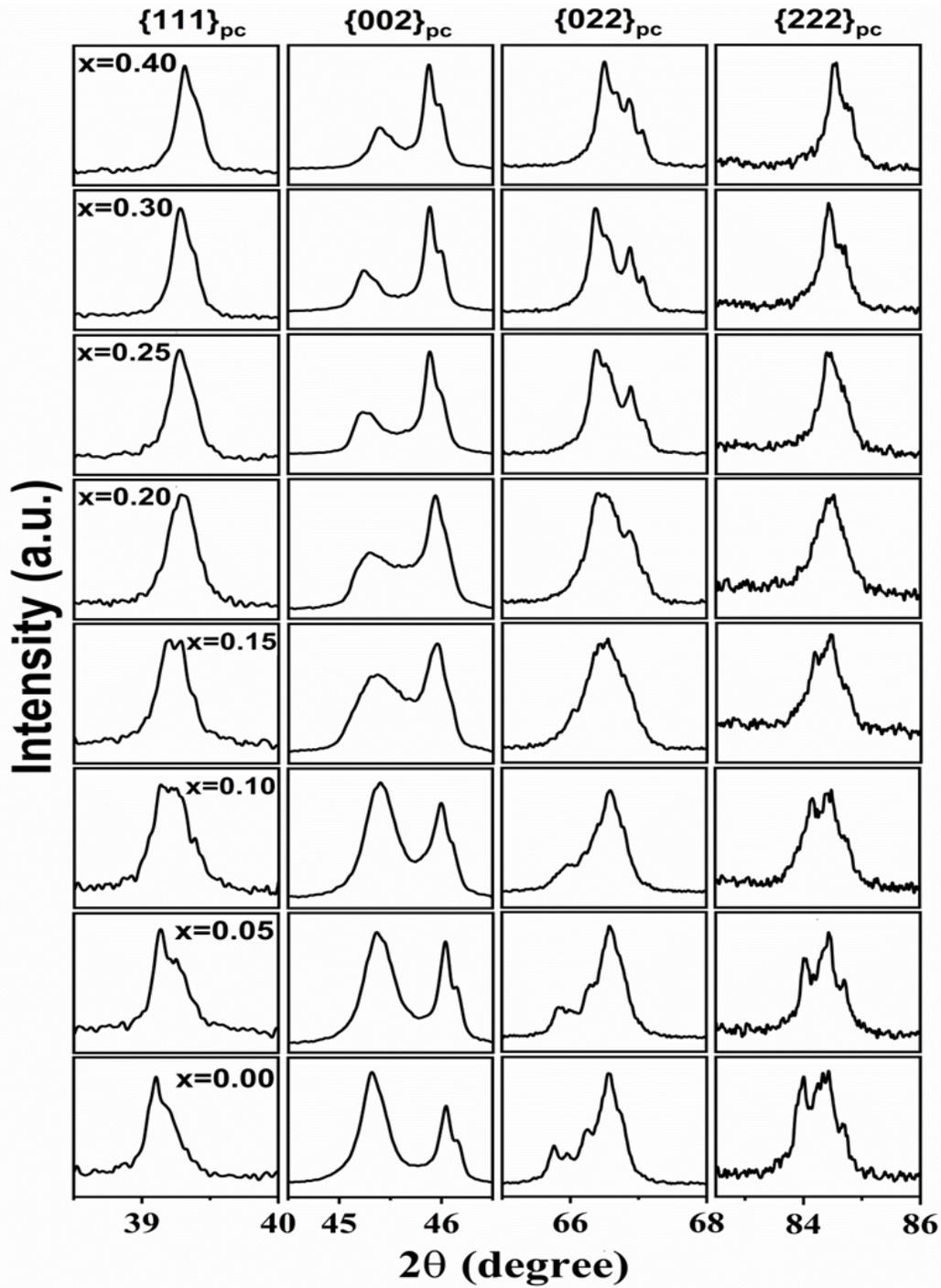

**Figure 1.** Evolution of XRD profiles of $\{111\}_{pc}$, $\{200\}_{pc}$, $\{220\}_{pc}$, and $\{222\}_{pc}$ pseudocubic reflections of KNLNT-$x$ ceramics with $0 \leq x \leq 0.4$ at room temperature.

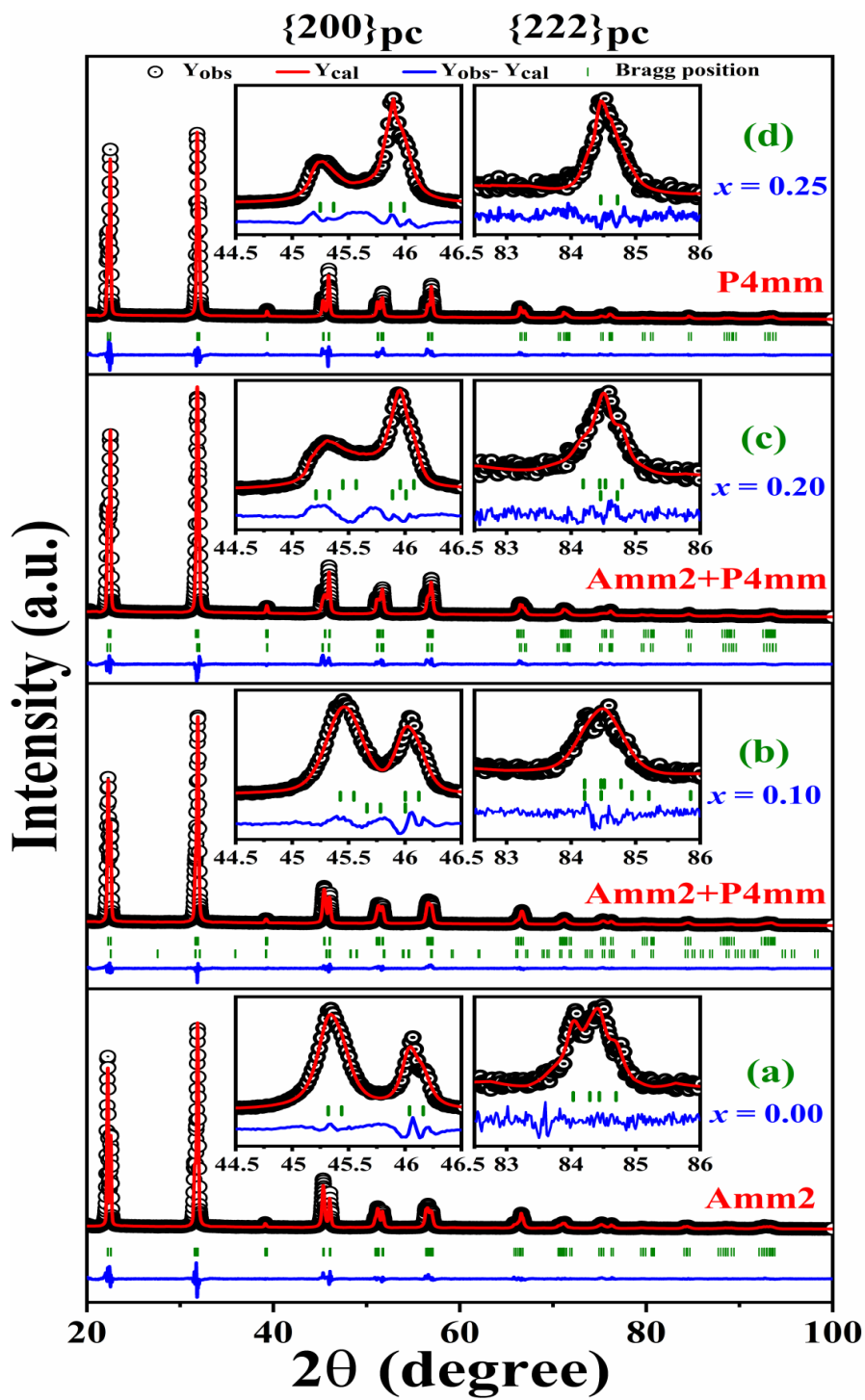

**Figure 2.** Rietveld refined x-ray powder diffraction patterns of KNLNT-*x* ceramics for **(a)** *x* = 0 using *Amm2* model **(b)** *x* = 0.10 & **(c)** *x* = 0.20 using *Amm2+P4mm* model **(d)** *x* = 0.25 using *P4mm* model.

## III. RESULT
### A. X-Ray Diffraction

Figure S1 shows the room temperature XRD patterns of KNLNT ceramics for different Ta concentrations. The presence of sharp and well-defined peaks (without any secondary phases) in the XRD patterns clearly suggest the formation of the perovkite structure in the ceramics. However, for the highest composition i.e. for $x = 0.4$, minor secondary peaks appear suggesting we have exceeded the Ta solubility limit.

In order to study the effect of Ta substitution on the crystal structure of KNLNT ceramics, firstly we have examined the splitting of pseudo-cubic $\{h00\}_{pc}$, $\{hh0\}_{pc}$ and $\{hhh\}_{pc}$ reflections, which is the best approach to identify the presence of different crystallographic phases such as cubic, tetragonal, orthorhombic and rhombohedral qualitatively. It has been reported that the pseudo-cubic reflections for the cubic symmetry are all singlet, whereas the tetragonal symmetry is characterized by the splitting of the $\{h00\}_{pc}$, and $\{hh0\}_{pc}$ reflections into doublet and singlet nature of the $\{hhh\}_{pc}$ reflection [5, 9, 10, 39]. The singlet nature of $\{hh0\}_{pc}$ and $\{h00\}_{pc}$ reflections and doublet nature of $\{hhh\}_{pc}$ reflection on the other hand, are characteristic of rhombohedral symmetry. However, for orthorhombic crystal structure, the $\{h00\}_{pc}$, $\{hh0\}_{pc}$, and $\{hhh\}_{pc}$ reflections are all doublets [10, 39, 40].

To better illustrate the phase evolution due to the splitting of pc reflections (i.e., to analyze the crystal structure), the magnified XRD patterns for the pseudo-cubic $\{111\}_{pc}$, $\{200\}_{pc}$, $\{220\}_{pc}$, and $\{222\}_{pc}$ reflections of KNLNT ceramics as a function of Ta concentration are shown in Fig. 1. In the XRD profiles, the peak splitting is minimal at lower-order reflections (lower 2θ), but more pronounced at higher-order reflections. Therefore, here we have concentrated on the $\{200\}_{pc}$, $\{220\}_{pc}$, and $\{222\}_{pc}$ higher order reflections to visualize the weak distortion. For $x = 0$, the $\{200\}_{pc}$ reflection observed around 45°, splits into a doublet (the (200) peak occurring at a lower Bragg's angle (2θ)) with the relative intensity ratio $I_{200}/I_{020}$ is about 2:1. Further, the $\{220\}_{pc}$ reflection exhibits an opposite tendency in the peak splitting compared to $\{200\}_{pc}$ reflection, and the $\{222\}_{pc}$ reflection splits into a doublet with nearly equal intensity. Also, the demanding criteria for determination of the single phase material is the resolution of the {222} reflection into their respective components [40]. These are the characteristics of a single-phase orthorhombic structure with *Amm2* space group and hence pure KNN ($x = 0$) crystallizes in orthorhombic symmetry.

A similar trend in the XRD profile is also observed for $x = 0.05$, suggesting orthorhombic structure of the unit cell. As the Ta concentration increases to $x = 0.10$, the shapes of the XRD profiles undergo a noticeable change in the profile shapes. These changes include a decrease in the splitting of the $\{200\}_{pc}$, $\{220\}_{pc}$, and $\{222\}_{pc}$ reflections and broadening of the peaks. As $x$ increases further ($x \geq 0.15$), an opposite trend in the splitting of the $\{200\}_{pc}$ reflection appear, where the (002) peak emerges with enhanced intensity as compared to the (200) peak. For $x = 0.15$ and 0.20, the splitting of $\{220\}_{pc}$, and $\{222\}_{pc}$ reflections further decrease and appear like a broad singlet. The broadening of the $\{200\}_{pc}$, $\{220\}_{pc}$, and $\{222\}_{pc}$ reflections suggest that there might be another crystallographic phase along with the orthorhombic phase due to the overlapping of two peaks. The opposite trend of the splitting of $\{200\}_{pc}$ reflection (compared to the $\{200\}_{pc}$ reflection of $x = 0.10$) and singlet nature of the $\{222\}_{pc}$ reflection strongly suggests the presence of a tetragonal phase along with the orthorhombic phase. For $x \geq 0.25$, the (002) peak becomes more pronounced and intense compared to the (200) peak and the $\{222\}_{pc}$ becomes a singlet. Concurrently, $\{220\}_{pc}$ reflection becomes a doublet with the stronger peak occurring at the lower angle side. The splitting of $\{200\}_{pc}$ reflection into (200), and (002) in a 1:2 intensity ratio and the singlet nature of the $\{222\}_{pc}$ reflection clearly suggests the tetragonal structure for $x \geq 0.25$. Therefore, visual observation of the profile pattern suggests orthorhombic structure for $x = 0$ and 0.05. With increasing Ta concentration, an additional tetragonal phase appears along with the orthorhombic phase up to $x = 0.20$. For $x \geq 0.25$, the doublet nature of $\{220\}_{pc}$ reflection and singlet nature of the $\{222\}_{pc}$ reflection clearly suggest a single-phase tetragonal structure. The distinct change in the diffraction pattern therefore suggests a composition-induced structural phase transition from orthorhombic to tetragonal structure through the phase co-existance region.

The qualitative analysis only provides a preliminary insight into the possible crystallographic phases present in the KNLNT ceramics. In order to confirm the exact crystal structure and the nature of the phase transitions (quantitatively) upon Ta substitution, the Rietveld refinement analysis has been carried out using the FULLPROF software [41]. Here the starting models for the refinement are proposed based on the predictions obtained from the qualitative visual study. In the refinement process, the pseudo-Voigt function has been used to model the peak shape while the linear interpolation technique is used to refine the background shape. Various refined parameters such as zero correction, scale factor, lattice parameters, Wyckoff positions, background, and half-width parameters (U, V, and W) are refined while keeping the occupancy fixed during the refinement. Here the black open circle, red line, blue line, and green scattered vertical lines represent the experimental data, theoretical data, the difference between theoretical and experimental data, and Bragg's peak positions, respectively.

Figure 2 shows the Rietveld-fitted XRD patterns of KNLNT ceramics at selected compositions ($x = 0$, 0.10, 0.20, and 0.25). To give a better illustration of our refinement, we have also shown the magnified view of the fitted data of the higher angle reflections $\{200\}_{pc}$ and $\{222\}_{pc}$ in the inset of Fig. 2. The refined data of the remaining compositions are given in the supplementary section (Figure S2). For $x = 0$ and 0.05, the XRD data fitted using the Rietveld refinement method with a single-phase *Amm2* crystal structure as shown in Fig. 2(a) and Figure S2(a). A satisfactory fit between the experimental and the theoretical pattern indicates that the sample also crystallizes in single-phase orthorhombic structure.

However, for $x = 0.10$, the Rietveld refinement with a single-phase *Amm2* crystal structure did not yield a satisfactory match between the theoretical and experimental patterns as shown in Figure S3(a). A dramatic improvement of the peak fitting occurs when we include the tetragonal phase along with the orthorhombic phase (*Amm2+P4mm*) as shown in Figure 2(b) and Figure S3(b). Further, the Rietveld refinement for $x = 0.15$ and 0.20 along with $x = 0.10$ are performed with the *Amm2+P4mm* coexistence model and the results show good agreement between the experimental and the suggested structural model as shown in Figure 2(b,c) and Figure S2(b). Based on the visual examination of the doublet nature of $\{200\}_{pc}$ and $\{220\}_{pc}$ reflections and singlet nature of $\{222\}_{pc}$ reflections, the structure should be tetragonal structure with *P4mm* space group. So we have fitted the XRD data for $x = 0.25$, using both dual phase of *Amm2+P4mm* and single phase *P4mm* model as shown in Figure S4. The single phase *P4mm* model resulted in a better fit between the experimental and chosen theoretical model. So for $x = 0.30$, and 0.40, we have also fitted with single phase *P4mm* model and the result shows a good fit. The refined structural parameters have been obtained using Rietveld refinement for all the samples are listed in Table S2 in the Supplementary Material. Thus, summarily, the Rietveld refinement on the XRD analysis reveals that KNLNT ceramic exhibits the single phase *Amm2* crystal structure for $x = 0, 0.05$, the coexistence of *Amm2+P4mm* structure for the composition range $0.01 \leq x \leq 0.20$, and tetragonal *P4mm* structure for $x \geq 0.25$. The phase transition from O to T through O+T coexistence phase can be represented by diffusion of phase transformation between O to T phase as well as the composition inhomogeneity, which is inevitable for the ceramic oxides [17].

### B. Raman Spectroscopy

Raman spectroscopy is very sensitive to crystal structure distortion and local heterogeneities that changes intensity, damping constant and frequency of Raman modes [42]. So, the effects of Ta on the structure and local heterogeneities of KLNN have been discussed by Raman spectroscopy. To dodge Bose-Einstein phonon population effect, the reduced Raman intensity, $I^r(\omega)$ was determined from the observed Raman scattering intensity, $I(\omega)$ as follows [43, 44].

$$I^r(\omega) = \frac{I(\omega)}{\omega[n(\omega)+1]} \quad \dots\dots\dots\dots\dots\dots\dots\dots\dots\dots\dots\dots\dots\dots\dots\dots\dots(1)$$

where, $n(\omega) = \frac{1}{\exp\left(\frac{\hbar\omega}{k_B T}\right)-1}$ is the Bose-Einstein population factor. $k_B$ is the Boltzmann constant and $\hbar$ is the Dirac constant. The calculated reduced Raman scattering spectra from the observed spectra is shown in Fig. 3(a). The reduced Raman spectra were fitted by a combination of a Lorentzian central peak (CP) and damped harmonic oscillators (DHO) in the frequency range 15~1000 to apprehend the Ta doping effects on KNLN [43, 44]:

$$I^r(\omega) = \frac{2A_{CP}}{\pi}\frac{\Gamma_{CP}}{4\omega^2+\Gamma_{CP}^2} + \sum_i \frac{A_i \Gamma_i \omega_i^2}{(\omega^2-\omega_i^2)^2+\omega^2\Gamma_i^2} \quad \dots\dots\dots\dots\dots\dots (2)$$

Here, $A_{CP}$ and $\Gamma_{CP}$ denote intensity and FWHM of the CP, respectively. $A_i$, $\Gamma_i$, and $\omega_i$ are the intensity, damping constant, and frequency of the $i^{th}$ optical Raman active mode, respectively.

The composition-dependent Raman scattering spectra of the KNLNT-$x$ recorded at room temperature is displayed in Fig. 3(a). At room temperature, the orthorhombic phase with *Amm2* symmetry of the KNLNT-$x$ ($x = 0.0$) is observed by Rietveld refinements of the XRD spectrum.

The irreducible representations of the orthorhombic phase with *Amm2* symmetry are $4A_1+A_2+4B_1+3B_2$ in accordance with group theory analysis [45, 46]. The direct symmetry assignment of vibrational modes of polycrystalline ceramic is difficult from the Raman scattering spectra; thus, we pursue the symmetry of the vibrational modes of the [$(K_{0.56}Na_{0.44})(Nb_{0.65}Ta_{0.35})O_3$, KNNT] single crystals [45].

As can be seen in Fig. (3b), the observed Raman scattering spectrum of the KNLNT-*x* (*x* = 0) of the KNN consist of mainly $B_2(TO_1)$ (~ 52 cm$^{-1}$), $B_1(TO_1)$ (~71 cm$^{-1}$), $A_1(TO_1)$ (~102 cm$^{-1}$), $A_1(TO_2)$ (~139 cm$^{-1}$), $A_1(LO_1)$ (~194 cm$^{-1}$), $B_1(TO_2)$ (~237 cm$^{-1}$), $A_1(TO_3)$ (~263 cm$^{-1}$), $A_1(LO_2)$ (~448 cm$^{-1}$), $B_2(TO_2)$ (~552 cm$^{-1}$), $A_1(TO_4)$ (~616 cm$^{-1}$), and $A_1(LO_3)$ (~860 cm$^{-1}$) in the frequency range of 15~1000 cm$^{-1}$. The broad weak mode near 715 cm$^{-1}$ (Figs. 3a and 3d) may be attributed to lattice disorder, which is induced by the ionic radii mismatch at crystallographic equivalent sites [9, 47]. The results are similar to those observed in orthorhombic ($K_{0.5}Na_{0.5}$)NbO$_3$ (KNN) [9]. Thus, the observed Raman modes reveal that KNLNT at *x* = 0.0 belongs to the orthorhombic phase [9, 45, 46]. This result is supported by dielectric and XRD results as well. The splitting of $A_1(TO_3)$, $B_2(TO_2)$, $A_1(TO_4)$, and $A_1(LO_3)$ modes is found and marked as $A_1(TO_3)_1$ (~263 cm$^{-1}$) and $A_1(TO_3)_2$ (~272 cm$^{-1}$), $B_2(TO_2)_1$ (~547 cm$^{-1}$) and $B_2(TO_2)_2$ (~564 cm$^{-1}$), $A_1(TO_4)_1$ (~595 cm$^{-1}$) and $A_1(TO_4)_2$ (~618 cm$^{-1}$), $A_1(LO_3)_1$ (~854 cm$^{-1}$) and $A_1(LO_3)_2$ (~863 cm$^{-1}$), respectively. The Raman modes splitting in KNLNT-*x* may be due to the different local order regions [9, 48]. It is observed in Fig. 3 (b, d) that the $B_1(TO_1)$ and $A_1(TO_2)$ modes disappear at *x* = 0.15, indicating the structural transition of the KNLNT-*x* from orthorhombic phase. Note that an over damped Raman $E(TO_1)$ mode appears at about 99 cm$^{-1}$ and $B_2(TO_2)_2$ mode disappear at *x* = 0.20 (Fig. 3b), which may correspond to the tetragonal phase of the KNLNT-*x* [9]. The tetragonal phase of the KNLNT at 0.20 <*x*≤ 0.25) with *P4mm* symmetry is confirmed by Rietveld refinement of the XRD spectrum. It is significant that the $B_1(TO_1)$ and $A_1(TO_2)$ modes disappear at *x* = 0.15, while $B_2(TO_1)$ mode corresponding to the orthorhombic phase persists, denoting structural transition from orthorhombic to mixed (orthorhombic and tetragonal) phase of the KNLNT-*x*. The coexistence of tetragonal (*P4mm*) and orthorhombic (*Amm2*) phases of KNLNT-*x* (0.10 <*x*≤ 0.25) is found by Rietveld refinement of the XRD spectra. However, the presence of prominent $B_1(TO_1)$ mode ssuggests that the properties of KNLNT at *x* = 0.15 may be dominated by the orthorhombic phase, which is supported by the phase fraction calculation as well. To understand the phase transition of KNLNT-*x*, we have studied composition-dependent frequency shift and FWHM of the prominent $A_1(TO_3)_1$ mode as well. As can be seen in the middle part of Fig. 3 (d), the frequency of the $A_1(TO_3)_1$ mode decreases while the FWHM increases continuously with Ta content except in the range of 0.10< *x*< 0.20. Note that the discontinuity of frequency and FWHM of $A_1(TO_3)_1$ mode the occurs at *x* = 0.15, indicating the structural phase transition of the KNLNT. Hence, the anomalous change of the frequency and FWHM of the $A_1(TO_3)_1$ mode may be due to the coexistence of both orthorhombic and tetragonal phase of KNLNT-*x* in the range of 0.10< *x*< 0.20 [5].

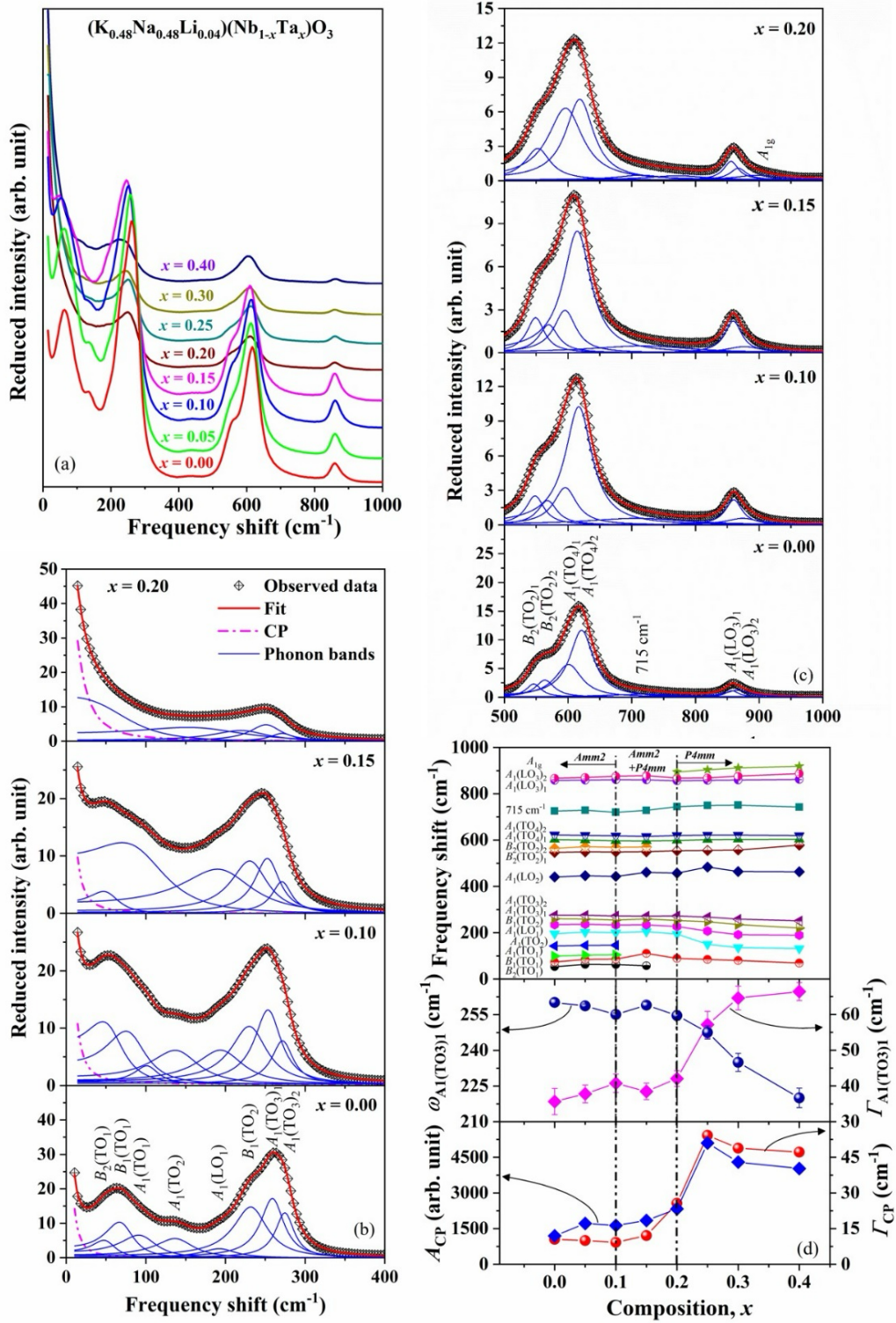

Figure 3: (Color online) (a) The composition dependent Raman spectra of KNLNT-$x$. (b,c) The fitted Raman spectra at some selected composition using Eq. (2). (d) The composition dependent frequency shift (upper part) of Raman active modes, frequency shift and FWHM of the $A_1(TO_3)_1$

mode (middle part), and the intensity (A) and FWHM (Γ) of the central peak (CP) (lower part) of KNLNT-$x$

The KNLNT-$x$ Raman spectra can also be rationalized by the translational modes of $K^+/Na^+/Li^{1+}$ cations and the internal modes of $NbO_6/TaO_6$ octahedrons. The $NbO_6/TaO_6$ octahedrons vibrational modes consists of $A_{1g}(v_1)+E_g(v_2)+2F_{1u}(v_3, v_4)+F_{2g}(v_5)+F_{2u}(v_6)$. The Raman modes $F_{1u}(v_3)$, $E_g(v_2)$ and $A_{1g}(v_1)$ are stretching and the rest of the modes are bending [45, 46]. In the low frequency range below 200 cm$^{-1}$, the Raman modes can be assigned to the rotation of the octahedron and the translational modes of the $K^+/Na^+/L^+$ cations. However, in the high frequency range of 200~1000 cm$^{-1}$, the Raman modes appear due to another internal vibrational octahedron. The $v_6$ mode near 139 cm$^{-1}$ related to $NbO_6/TaO_6$ octahedron may appear in the low frequency range as well [46]. The mode at about 194 cm$^{-1}$ is associated to $K^+/Na^+/Li^+$ cations versus $NbO_6/TaO_6$ octahedron, whereas the modes near 52 cm$^{-1}$ and 71 cm$^{-1}$ are related to the translational modes of $K^+/Na^+/Li^+$ cations [46]. The rotational mode of the $NbO_6/TaO_6$ octahedron appear near 102 cm$^{-1}$ [45]. The peaks at about 237 cm$^{-1}$ and 263 cm$^{-1}$ are marked as $v_5$ mode. The $v_4$, $v_2$, $v_1$, and $v_3$ modes of the $NbO_6/TiO_6$ octahedron appear near 448 cm$^{-1}$, 547 cm$^{-1}$, 618 cm$^{-1}$, and 715 cm$^{-1}$, respectively [46]. The peak at about 860 cm$^{-1}$ is identified as the coupled $v_1+v_5$ mode [46]. Note that the intense $v_5$ (263 cm$^{-1}$) and $v_1$ (618 cm$^{-1}$) modes become weak upon increasing Ta content in KNLNT [Fig. 3a]. The presence of strong $v_1$ and $v_5$ modes suggest the near-perfect $NbO_6/TaO_6$ octahedron of the orthorhombic phase KNLNT with space group *Amm2* at $x = 0$ [9]. Note that $v_5$ (263 cm$^{-1}$) and $v_1$ and modes of the KNLNT at $x = 0$ splits into two peaks. The mode splitting is attributed to the substitution of Nb ions with Ta ions, which breaks the symmetry of the $NbO_6/TaO_6$ octahedron due to the distortion of crystal structure [9, 45]. These modes redshift with increasing the Ta content (middle part of Fig. 3 (d)), indicating the substitution of Nb ions with Ta ions may weaken the binding strength of the octahedron giving rise to lengthening the distance between Nb/Ta cations and its coordinated oxygen due to their lattice mismatch [9]. The octahedral breathing $A_{1g}$ mode appears near 894 cm$^{-1}$ in the tetragonal phase of KNLNT-$x$ (0.20 ≤$x$≤ 0.40). The octahedral breathing $A_{1g}$ mode is also observed in the tetragonal phase of the BZT-$x$BCT system [5]. The symmetric $A_{1g}$ mode is Raman inactive and does not result in polarization change [49], whereas the asymmetric $A_{1g}$ mode becomes Raman active because of the presence of dissimilar ions in the centre of the octahedral.

To understand the nature of phase transition and Ta doping effects on local polar clusters in KNLNT-$x$, the FWHM of the CP, which is inversely proportional to the relaxation time of precursor dynamics [50], has been studied as a function of composition. The existence of the CP and soft mode phenomena is evidence of the order-disorder and the displacive nature of the ferroelectric phase transition, respectively. In the present study, the low-frequency $B_2(TO_1)$ mode at about 52 cm$^{-1}$ is observed in the orthorhombic phase at $x = 0$, whereas an overdamped $E(TO_1)$ mode at about 89 cm$^{-1}$ is attributed to the tetragonal phase of at 0.20 ≤$x$≤ 0.40. It is challenging to opine on the soft-mode behaviour of $B_2(TO_1)$ and $E(TO_1)$ modes due to the lack of temperature dependent Raman scattering results in the present study. Therefore, the existence of the prominent CP can indicate the order-disorder behavior of ferroelectric phase transition of KNLNT-$x$ [43,44,50]. In ferroelectric materials, the dynamic polar clusters called polar nano regions (PNRs) begin to develop at the Burns temperature ($T_B$) [51], and these dynamic PNRs grow into static PNRs at an intermediate temperature ($T^*$) [43,44,50]. These static PNRs turn into nano-domain states in the ferroelectric phase that are randomly oriented, and transform into macro-domain states owing to local polarization freezing [10, 52]. It is found from lower part of Fig. 3(d) that the value of $\Gamma_{CP}$ decreases in the composition range of $0 \leq x \leq 0.10$, where KNLNT belongs to the orthorhombic phase, increases in the range of $0.10 < x \leq 0.25$ where KNLNT belongs to the mixed phase, and then decreases in the range of $0.25 < x \leq 0.40$ in which KNLNT belongs to the tetragonal phase, with increasing Ta content. This result suggests that the size and/or number density of oriented nano-domain states may increase with Ta content [5,50]. In the range of $0.10 < x \leq 0.25$, the correlation among nano-domain states may be weakened and/or broken owing to the coexistence of phases which increases $\Gamma_{CP}$.

### C. Microstructural studies

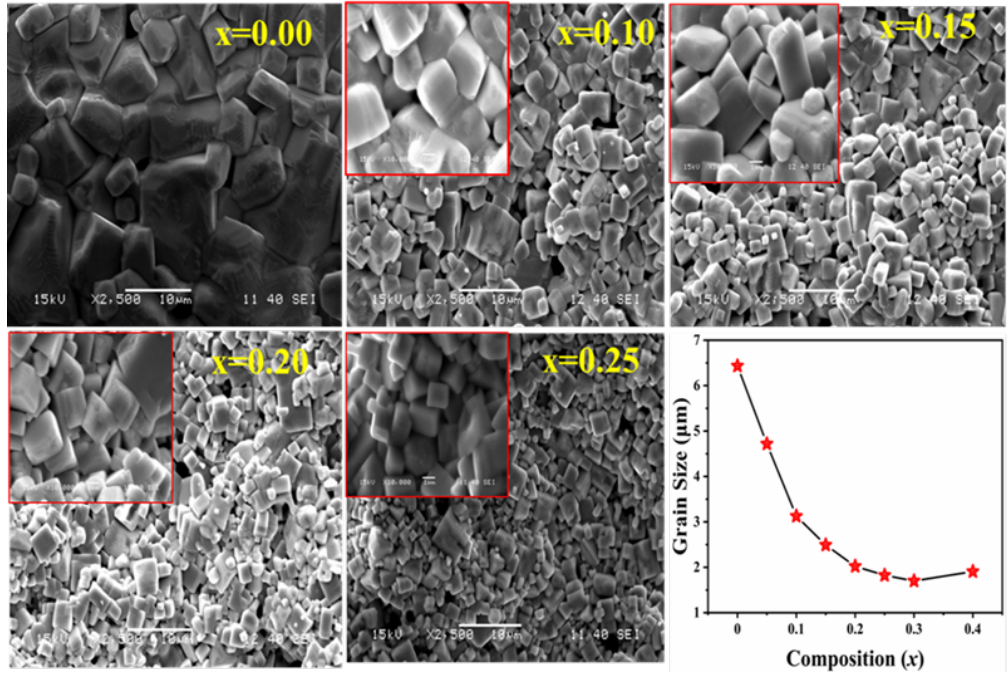

**Figure 4.** SEM micrographs of KNLNT-$x$ ceramics for (a) $x = 0$, (b) $x = 0.10$, (c) $x = 0.15$, (d) $x = 0.20$, (e) $x = 0.25$, (f) variation of grain size with composition.

Figure 4 shows the SEM micrographs of KNLNT-$x$ ceramics for $x$ = 0, 0.10, 0.15 0. 0.20, 0.25 respectively (other compositions are shown in the Supplementary Figure S5). The observed microstructure is significantly influenced by the variation of *x* in KNLNT-*x* systems. For all compositions, the grain sizes are calculated using ImageJ software and the average grain sizes are calculated from the Histogram. The histogram of the grain size distribution of some selected samples is also included in the supplementary file (Figure S6). For *x* = 0, the microstructure consists of different grain sizes (with larger grains as well as smaller grains) distributed throughout the microstructure. This is due to the abnormal grain growth process [19]. The grain size decreases and a compact microstructure emerges with an increase in Ta concentration. For higher concentrations, the micrograph consists of well-faceted and quite uniform distribution of fine grains. We have also plotted the variation of average grain sizes with Ta concentrations and a sharp decrease in the grain size at lower Ta concentration is observed. For higher Ta concentration ($0.1 \leq x \leq 0.4$), the grain size varies between 3 to 1.5 μm. The homogeneous distribution of fine grains is consistent with the enhanced physical properties described below [29].

## D. Dielectric Properties

The temperature-dependent dielectric permittivity ($\varepsilon_r$) and dielectric loss (tan$\delta$) at different frequencies were investigated from RT to 500 °C to study the phase transition behavior in KNLNT-*x* ($0 \leq x \leq 0.4$) ceramics. Figure 5 (a-e) shows the variation of $\varepsilon_r$ and tan$\delta$ as a function of temperature for *x* = 0.00, 0.10, 0.15, 0.20, and 0.25 at selected frequencies of 1 kHz, 5 kHz, 10 kHz, 50 kHz, and 100 kHz (the other compositions arewere shown in the Supplementary Figure S7). A decrease in dielectric permittivity with increasing frequency is observed for all the compositions, which indicates the polar dielectric behavior of the samples [5,53]. For *x* = 0, the temperature dependent $\varepsilon_r$ shows two distinct anomalies corresponding to ferroelectric phase transitions in the measured temperature range (Fig. 5 a). The broad dielectric anamoly observed around 140 °C corresponds to an orthorhombic-to-tetragonal polymorphic ($T_{O-T}$) phase transition, whereas the sharp dielectric maximum observed at 440 °C corresponds to the tetragonal-to-cubic ($T_C$) phase transition, i. e., ferroelectric to paraelectric phase transition [12, 31]. A similar behavior (i.e., two anamolies) is observed in the tan$\delta$ versus T graph for all the frequencies further conforming the ferroelectric phase transtions as shown in the inset of Fig. 5a. We have also prepared and measured the temperature-dependent $\varepsilon_r$ and tan$\delta$ of pure $K_{0.5}Na_{0.5}NbO_3$ [9]. The phase transition temperatures, i.e., $T_{O-T}$ and $T_C$ values for pure KNN were found to be at 210 °C and 405 °C, respectively. So, Li substituted KNN (KNLN, *x* = 0) shows an increased $T_C$ and a decreased value of $T_{O-T}$, which is consistent with literature [15]. On the other hand, when Ta is substituted into the KNLN sample, both the phase transition temperatures decrease and broaden (Figure 5, and Figure S7). For clear observation of the effect of Ta doping on the phase transition temperatures of KNLNT-*x* ceramics, the dielectric constant versus temperature for all the compositions are plotted at a constant frequency of 10 kHz as shown in Figure 5f. Both the transition peaks corresponding to $T_{O-T}$ and $T_C$ shift to lower temperature, and for *x* ≥ 0.20 we do not observe any anomaly corresponding to $T_{O-T}$ suggesting the shift of $T_{O-T}$ to below room temperature.

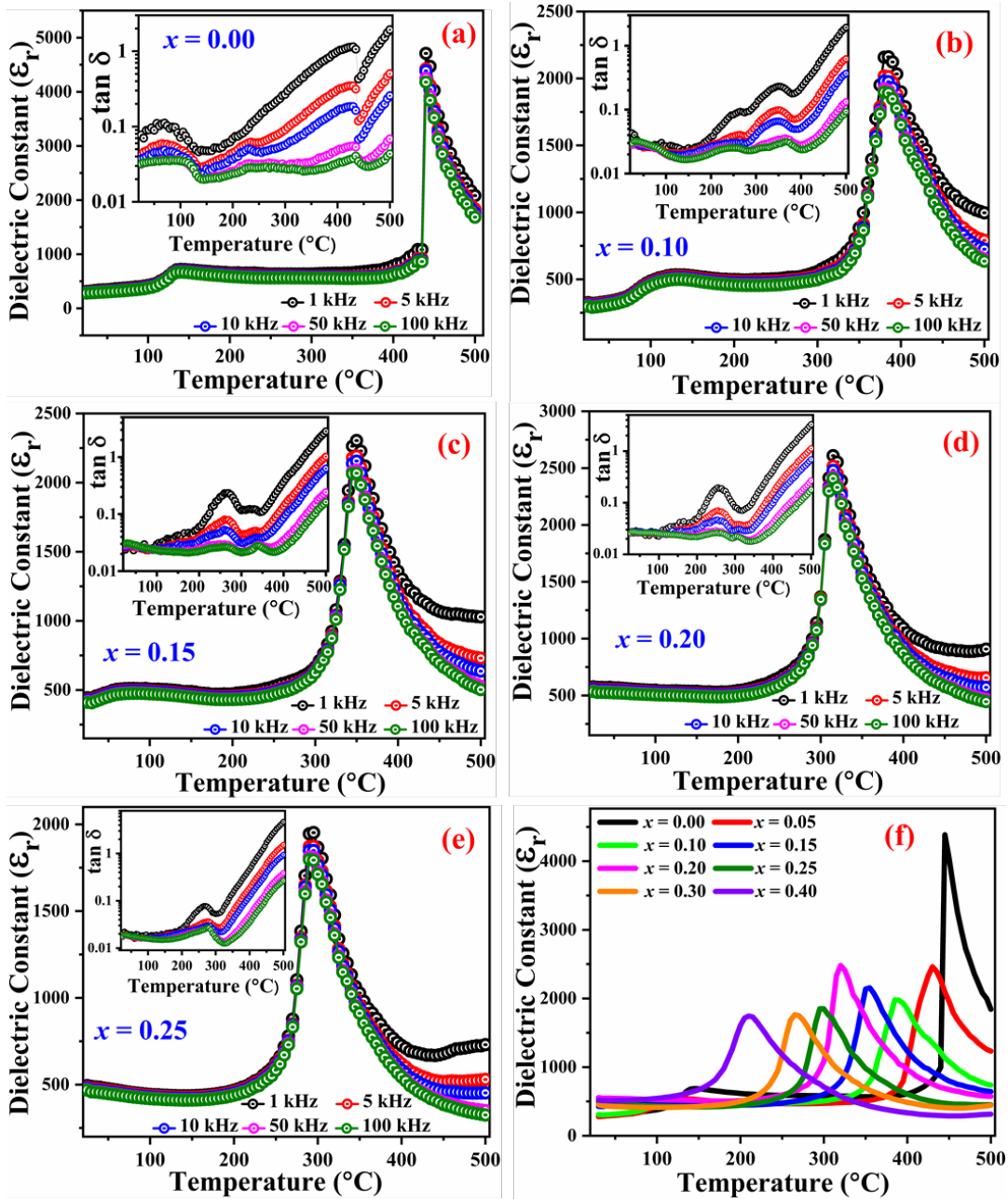

**Figure 5**. Variation of $\varepsilon_r$ and $\tan\delta$ as a function of temperature at selected frequencies of KNLNT-*x* ceramics for (a) *x* = 0, (b) *x* = 0.10, (c) *x* = 0.15, (d) *x* = 0.20, (e) *x* = 0.25, and (f) comparison of dielectric constant with temperature for all compositions at 10 kHz.

Note that both the phase transtions are broadened due to Ta substitutions and strongly depend on the compositions. In order to visualize the broadening of the orthorhombic-tetragonal transition peaks ($T_{O-T}$), we have enlarged the temperature-dependent dielectric constant plot around the $T_{O-T}$ region for *x* ≤ 0.15 which is shown in the supplementary Figure S8. For *x* = 0.00 (KNLN), the transition peak corresponding to $T_{O-T}$ is sharp and frequency-independent (Figure S8a). However, the $\varepsilon_r$ versus T plot for *x* = 0.05 shows slight broadening of this ferro-ferro phase transition peak compared to KNLN sample (Figure S8b). As the Ta concentration increases further ($x = 0.10$), the broadening of the $T_{O-T}$ peak increases (Figure S8c) and for higher concentration of Ta i.e. for $x = 0.15$, the phase transition peak becomes more broaden and shows a diffusive nature. The broadening of the phase transition ($T_{O-T}$) peak with increasing Ta concentration may be due to the increase in phase fraction of the tetragonal phase in the phase coexistence region. A similar observation has also been reported in multi-component substituted KNN system by Gao *et al.* [54] and Sm-doped KNN system by us also [55].

Further, to quantify the degree of diffuseness, temperature-dependent dielectric permittivity data in the paraelectric region have been fitted using modified Curie Wiess law [1,5]. The modified Curie-Weiss equation can be expressed as

$$\frac{1}{\varepsilon_r} - \frac{1}{\varepsilon_m} = \frac{(T-T_C)^\gamma}{C} \quad \ldots \ldots \ldots \ldots \ldots \ldots \ldots \ldots \ldots \ldots \ldots \ldots \ldots \ldots (3)$$

where $\varepsilon_m$ = highest value of $\varepsilon_r$ at $T_C$, $\gamma$ = degree of diffuseness, and $C$ = modified Curie Weiss constant. The value of $\gamma$, which ranges from 1 to 2, can be calculated from the slope of the $\ln(1/\varepsilon_r-1/\varepsilon_m)$ versus $\ln(T-T_m)$ plot. Normal ferroelectric materials are found to have a degree of diffuseness parameter of the order of 1, whereas very diffuse/relaxor-type ferroelectric materials have a degree of diffuseness of 2 [1, 5]. Further, the degree of diffuseness of KNLNT-$x$ ceramics for $0 \leq x \leq 0.40$ have been extracted from the graphs of $\ln(1/\varepsilon_r-1/\varepsilon_m)$ versus $\ln(T-T_m)$ at 10 kHz and are plotted in the paraelectric region as shown in Figure S9. The linear fit of the log-log plot suggests that the modified Curie-Weiss law has been verified. For $x = 0$, the value of $\gamma$ is found to be 1.21, and the $\gamma$ value increases with increasing Ta concentration, which is listed in the Supplementary Table S3. The rise in the values of $\gamma$ suggests that the diffuse nature of phase transition increases with increaseing Ta doping.

## E. Ferroelectric properties

The polarization versus electric field hysteresis loop (*P-E* loop) measurements were performed at room temperature to examine the ferroelectric characteristics of KNLNT-$x$ ceramics. The *P-E* loops of KNLNT-$x$ ceramics with $0.00 \leq x \leq 0.40$ were recorded at a frequency of 5 Hz at RT, as shown in Fig. 6. For all KNLNT-$x$ ceramics, the PE loop shows a well defined, nonlinear and saturated loop suggesting a good ferroelectric nature (shown in the inset of Figure 6) [5].

Further, the coercieve field ($E_C$) and remnant polarization ($2P_r$) were calculated from the *P-E* loops using the formula $2P_r^0 = P_r^{+0} - P_r^{-0}$ and $E_c^0 = (E_c^{+0} - E_c^{-0})/2$ for all the compositions [56]. However, with increasing Ta concentration, there is a systematic change in the behavior of the PE loops. The PE loop of KNLN ceramic shows higher $2P_r$ and $E_C$ values compared to the Ta modified samples. It has been reported by Shirane *et al.* that the remnant polarization ($2P_r$) and coercive field ($E_C$) are higher in the orthorhombic phase compared to the other crystallographic phases [57]. As KNLN crystallizes in the orthorhombic structure, this may be the reason for the observed higher ferroelectric properties ($2P_r$ and $E_C$) compared to the Ta- substituted samples.

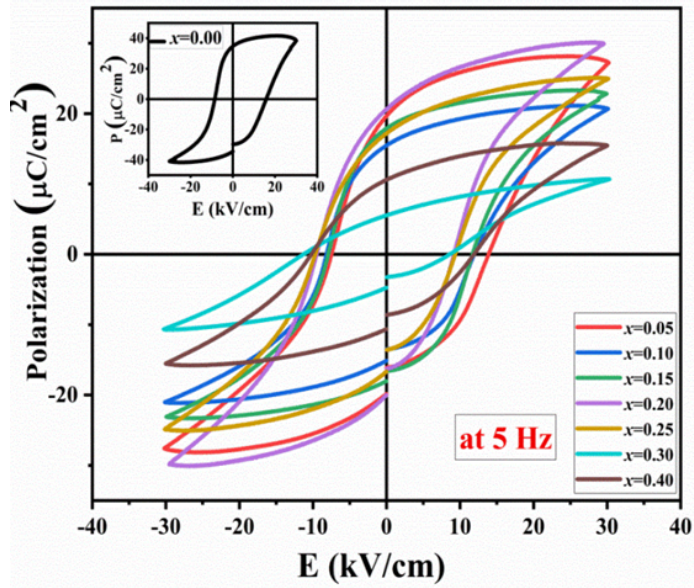

**Figure 6.** *P-E* hysteresis loop of KNLNT-*x* ceramics with $0 \leq x \leq 0.4$ at room temperature. The inset figure shows the P-E loop for *x* = 0.00.

Figure 7 shows the variation of (a) $2P_r$ and $E_C$ at RT, (b) dielectric constant and piezoelectric coefficient ($d_{33}$) at RT, (c) phase transition temperatures ($T_{O-T}$ and $T_C$) as a function of *x* for the KNLNT-*x* ceramics with $0 \leq x \leq 0.4$. For a better representation of the variation of physical properties with composition, we have divided the plot into three regions based on the crystal structure, (i) Region I for *x* = 0.00 to 0.10 (purely orthrorhombic strucutre), (ii) Region II for *x* = 0.10 to 0.20 (orthrorhombic+ tetragonal (O+T) phase co-existence region) and (i) Region III for *x* = 0.20 to 0.40 (purely tetragonal structure). In Region I, both the $2P_r$ and $E_C$ values steadily decline with an increase in Ta content. The $2P_r$ and $E_C$ values drop from 69.11 µC/cm² and 12.18 kV/cm for KNLN (*x* = 0.00) to 30.59 µC/cm² and 10.04 kV/cm for KNLNT-*x* (*x* = 0.10) respectively. Higher values of $2P_r$ and $E_C$ in purely orthorhombic phase is more in accordance with the literature [57]. The decrease of the $2P_r$ and $E_C$ value with *x* in the region I may be due to a decrease in the orthorhombic distortion. Both the dielectric constant ($\varepsilon_r$), and piezoelectric coefficient ($d_{33}$) measured at room temperature slightly increase with an increase in *x*.

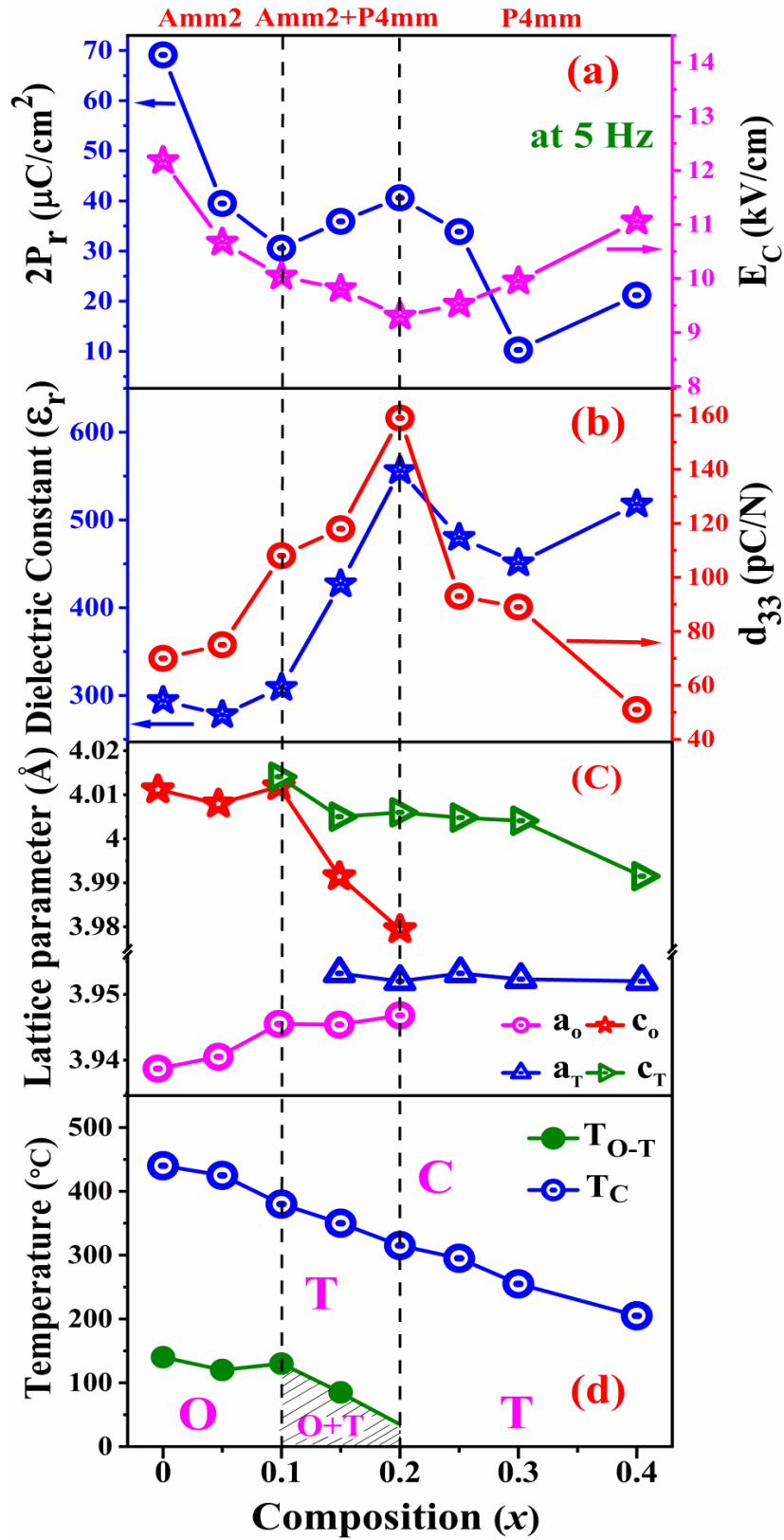

**Figure 7.** Variation of (a) ferroelectric parameters ($2P_r$ and $E_c$) at RT, (b) dielectric constant ($\varepsilon_r$) at 10 kHz and piezoelectric coefficient ($d_{33}$) at RT, (c) lattice parameters, and (d) phase transition temperatures ($T_{O-T}$ and $T_C$) with $x$ for KNLNT-$x$ ceramics.

In region II, the 2$P_r$ increases with increasing Ta concentration, and a maximum value of 40.62 µC/cm$^2$ is observed for $x = 0.20$. On the other hand, the coercive field $E_C$ decreases with increasing $x$ and is a minimum for $x = 0.20$ ( 9.3 kV/cm). The dielectric constant and piezoelectric coefficient increase with increasing $x$ and display a local maxima at $x = 0.20$. It may be noted that the rate of increase of both the dielectric constant and piezoelectric coefficient values is more in the phase co-existence region compared to the single-phase region. In Region III, the 2$P_r$ values decrease with increasing Ta composition. The coercive field, however increases with increasing Ta substitution. Both the dielectric constant and piezoelectric coefficient decrease with the increasing Ta concentration in this region. The variation of lattice parameters with composition ($x$) for KNLNT-$x$ is shown in Figure 7 (c). Furthermore, the phase diagram constructed from the RT XRD data, Raman spectra, and temperature-dependent dielectric properties is displayed in the Figure 7 (d). Both the phase transition temperatures decrease linearly with the increase in the Ta- concentrations. In region I, the phase transition sequence is orthorhombic to tetragonal then to cubic with increasing temperature. In the region II, the phase co-existence of O+T crystallographic phases, transform to tetragonal and finally to cubic as a function of increasing temperature. The transition from tetragonal to cubic is observed in region III with a rise in temperature.

## IV. DISCUSSION

The composition-driven structural phase transition in ferroelectric solid solutions where either a crystallographic phase with lower symmetry exists or two crystallographic phases having nearly equal phase fraction and nearly equal thermodynamic free energy co-exist for a small composition range is known as MPB [20, 58, 59]. Around the MPB region, there is abrupt enhancement of the dielectric, ferroelectric and piezoelectric properties observed. The classical PbZr$_{1-x}$Ti$_x$O$_3$ is a solid solution of PbZrO$_3$ (antiferroelectric with Rhombohedral crystal structure) and PbTiO$_3$ (ferroelectric with tetragonal crystal structure) and around the MPB composition, enhanced physical properties are observed [58, 59]. The MPB region in the case of PZT is independent of temperature but strongly depends on composition [58, 59].

As shown in the phase diagram (Fig. 7(d)), it has been observed that $T_{O-T}$ and $T_C$ shifts towards RT with an increase in Ta concentration and the phase boundary between the tetragonal and orthorhombic phase is not vertical. Unlike PZT, similar temperature-dependent MPB regions have been reported in many relaxor ferroelectric solid solutions such as PMN-PT, Pb(Zn$_{1/3}$Nb$_{2/3}$)O$_3$-PbTiO$_3$ (PZN-PT), Pb(Ni$_{1/3}$Nb$_{2/3}$)O$_3$-Pb(Zr,Ti)O$_3$ (PNN-PZT) [59, 30, 60]. So the phase (O+T) co-existence region of $0.1 \leq x \leq 0.2$ as observed in Fig. 7 may be due to the MPB or due to the shifting of $T_{O-T}$ towards room temperature for KNN-based system. Although phase co-esistence is the primary criteria for the enhancement of the piezoelectric and ferroelectric properties, we are interested in distinguishing if this is due to MPB or due to the shifting of the PPB towards RT.

It has also been reported that enhanced piezoelectric properties in the KNN system is due to well-faceted and uniform distribution of grains rather than a MPB [29, 61]. In the phase coexistence region ($0.1 \leq x \leq 0.2$) as well as in the tetragonal phase for higher Ta concentrations ($x$ = 0.3, and 0.4), the grains are of comparable sizes with similar distribution. Hence the microstructure is not associated with the enhancement of physical properties for our case.

In our case, the phase coexistence (low symmetry) region, ($0.1 \leq x \leq 0.2$) the phase coexistence (O+T) is separated by O (at low Ta concentration) and T (higher Ta concentration) with different symmetry as shown in Figure 6. For a MPB, the phase coexistence region can be 2 to 15 mol. % (e.g. $PbTiO_3$ in $Pb(Ti_{1-x}Zr_xO_3)$) depending upon the processing and substitution of cations [61, 62], and for our case it is ~10 mol.%. There is a discontinuous change in the lattice parameters (shown in Fig. 7 (c)) for $x$ = 0.10 and $x$ = 0.20 (starting and ending point of the phase coexistence region), which suggest a $1^{st}$ order phase transition and observed around MPB regions in case of ferroelectric systems [15, 62, 63, 64]. For $x$ = 0.2 (which is in the tetragonal side of the phase coexistence region), both the tetragonal and orthorhombic phases have nearly equal phase fractions which suggests they have thermodynamically similar energy [14, 29]. The availability of more polarization vectors (12 for orthorhombic and 6 for tetragonal) in the phase coexistence region is also expected [1, 65]. As the energy difference between O and T (tetragonal) phases is small in this region, the inter-conversion of both the O and T phases is easy, leading to easy polarization rotation [30, 66, 67]. For multi-component substituted KNN with different concentrations of different dopants, the change of crystal structure from O to T phase is accompanied by the structural heterogeneity [54]. In our case Ta- substitution increases the volume fraction of the tetragonal phase accompanied by induction of a large amount of structural heterogeneity. Thus, the local structural heterogeneity facilitates easy polarization rotation by flattening the polarization anisotropy energy, which synergistically contributes to the enhancement of the piezoelectric properties of the Li/Ta modified KNN system [54].

In the phase coexistence region II, there is an increase in the $2P_r$ values up to $x$ = 0.20 and a decrease in the $E_C$ value up to $x$ = 0.20 i.e., we have observed a maximum value of $2P_r$ and minimum value of $E_C$ for $x$ = 0.20. Above these concentrations, reverse trend is being observed. Iwota *et al.* based on the Landau Theory reported that the $E_C$ value becomes extremely small in the MPB region [68]. This decrease in the $E_C$ value may be due to the isotropic nature of the free energy function of the polarization component in the MPB composition [68, 69]. The sharp enhancement of the $2P_r$ value in the MPB is due to the enhancement of strain and ferroelasticity, which enhance the piezoelectric properties [69]. We have also observed a maximum value of $\varepsilon_r$ and a maximum $d_{33}$ value for $x$ = 0.20 at RT. So maximum value of all the physical properties has been observed for the critical composition $x$ = 0.20. Although, for $x$ = 0.20 (where the $T_{O-T}$ is close to RT), we have observed maximum piezoelectric properties, but the rate of change of the piezoelectric properties is more in the phase coexistence region [21, 5]. So in our case, the enhancement of the physical properties in the phase co-existence region is due to the MPB rather than shifting of the PPB towards RT.

For the ferroelectric perovskite oxides, the piezoelectric coefficient $d_{33}$ can be written as

$$d_{33} = 2Q_{11}\varepsilon_0\varepsilon_{33}P_3 \dots \dots \dots \dots \dots \dots \dots \dots \dots \dots \dots \dots \dots \dots \dots \dots \dots \dots \dots \dots (4)$$

where $P_3$ = polarization along the polar axis and approximately equals to remnant polarization ($P_r$) in this case, $\varepsilon_0$ = dielectric permittivity in free space, $\varepsilon_{33} = \varepsilon_r$ is the dielectric permittivity and $Q_{11}$ = electrostrictive coefficient which varies between 0.05 and 0.1 m$^4$ C$^{-2}$ [14, 70, 71]. Therefore, the enhancement of $d_{33}$ is directly related to $\varepsilon_r$ and $P_r$. As we have observed a maximum dielectric and ferroelectric polarization for $x = 0.2$, it is expected that the maximum piezo-electric coefficient is observed for that critical composition $x = 0.2$.

Although, the phase coexistence region is $0.1 \leq x \leq 0.2$, *why the maximum dielectric, ferroelectric, and piezoelectric properties are observed for $x = 0.2$?*

It has been reported that extrinsic contributions like domain switching in ferroelectrics are being contributed more in the phase co-existence region as compared to the single phase so that significant strain is being developed. So we expect that in the phase co-existence region, the domain switching induces significant strain, which further flattens the polarization anisotropy energy and strengthens the electro-mechanical coupling [5]. Based on *in situ* XRD experiments, Pramanik *et al.* correlated a decrease in the tetragonality with an enhancement of the domain wall motion [72]. We have calculated the tetragonality [tetragonality = $(c/a - 1)$] of both the tetragonal and orthorhombic phases as shown in Figure S 10 [5, 72]. A minimum tetrgonality for the orthorhombic phase is observed for $x = 0.2$ and the tetrgonality of tetrogonal phase is constant in the phase coexistance region, which is responsible for enhanced domain wall motion. The domain wall motion is also well corroborated with the CP behavior in the Raman analysis. Thus the enhancement of the piezoelectric and ferroelectric properties for the critical composition $x = 0.2$ around the MPB regions is due to (1) easy polarization rotation, and (2) the maximum domain wall motion due to flattening the curvature of the free energy profile.

## V. CONCLUSIONS

In summary, high density KNLNT-*x* ceramics were successfully fabricated by the conventional solid state reaction route. The phase transition behavior, surface morphology, dielectric, ferroelectric and piezoelectric properties have been systematically investigated over a wide range of experimental conditions. The XRD analysis using Rietveld refinement suggest a composition-driven structural phase transition from the orthorhombic (*Amm2*) phase to the orthorhombic+tetragonal (*Amm2+P4mm*) dual phase and finally to tetragonal phase at room temperature, which is also confirmed from the Raman spectroscopic analysis. The temperature-dependent dielectric properties suggest that, with increase in Ta concentration, both the phase transition temperatures ($T_{O-T}$ and $T_C$) decrease systematically along with the broadening of the transition peaks and for higher composition of Ta ($x > 0.15$), the $T_{O-T}$ shifts below room temperature. A phase diagram has been constructed based on the phase transtions behvaiour obtained from XRD, Raman and dielectric results. A maximum value of ferroelectric and piezoelectric properties around the MPB has been observed for the critical composition $x = 0.20$. The enhancement of piezoelectric properties is due to the existence of MPB rather than PPB governed by easy polarization rotation, and the maximum domain wall motion.

**ACKNOWLEDEMENT:**

Satyaranjan Sahoo acknowledges the Ministry of Human Resource Development, India for the institute research fellowship. We acknowledge Dr Banarji Behera School of Physics, Sambalpur University, Jyoti Vihar, Burla for providing experimental facility and fruitful discussion. We also acknowledge the partial financial support from IUAC, New Delhi, India, under the grant (IUAC/XIII.7/UFR-68310).

**DATA AVAILABILITY**

The data supporting the findings of the present study are available to the readers from the corresponding author upon reasonable request.

# Supporting Information

# Morphotropic Phase Boundary (MPB) Induced Enhancement of Ferroelectric and Piezoelectric Properties in Li and Ta modified $K_{0.5}Na_{0.5}NbO_3$

Satyaranjan Sahoo[1], Dhiren K. Pradhan[2], Shalini Kumari[3], Abhisikta Sahu[1], Koyal Suman Samantaray[4], Vikas N. Thakur[5,6], Anupam Mishra[7], Md. Mijanur Rahaman[8], Ashok Kumar[5,6], Reji Thomas[9,10], Philip D. Rack[2], and Dillip K. Pradhan[1*]

[1]Department of Physics and Astronomy, National Institute of Technology Rourkela, Rourkela, Odisha 769008, India

[2] Department of Electrical and Systems Engineering, University of Pennsylvania, Philadelphia, PA, 19104, USA.

[3]Department of Materials Science & Engineering, The Pennsylvania State University, University Park, Pennsylvania 16802, USA

[4]Department of Physics, Indian Institute of Technology Indore, Indore, 453552, India

[5]CSIR-National Physical Laboratory, Dr. K. S. Krishnan Marg, New Delhi 110012, India

[6]Academy of Scientific and Innovative Research (AcSIR), Ghaziabad 201002, India

[7]Department of Ceramic Engineering, National Institute of Technology Rourkela, Rourkela, Odisha 769008, India

[8]Department of Materials Science and Engineering, University of Rajshahi, Rajshahi 6205, Bangladesh

[9]Division of Research and Development, Lovely Professional University, Jalandhar-Delhi G.T. Road, Phagwara, Punjab 144411, India

[10]School of Chemical Engineering and Physical Sciences, Lovely Professional University, Jalandhar-Delhi G.T. Road, Phagwara, Punjab 144411, India

[*] Author to whom correspondence should be addressed: dillip.pradhan79@gmail.com, pdhiren@seas.upenn.edu

# Figure Captions (Supplementary Material)

**Figure S1.** Room temperature XRD patterns of KNLNT-$x$ with $0 \leq x \leq 0.4$.

**Figure S2.** Rietveld refined x-ray powder diffraction patterns of KNLNT-$x$ for **(a)** $x = 0.05$ using *Amm2* model, **(b)** $x = 0.15$ using *Amm2+P4mm* model, **(c)** $x = 0.30$ and **(d)** $x = 0.40$ using *P4mm* model.

**Figure S3.** Rietveld refined x-ray powder diffraction patterns of KNLNT-$x$ for $x = 0.10$ using **(a)** *Amm2* and, **(b)** *Amm2+P4mm* model. The inset shows magnified view of the fitted $\{200\}_{pc}$ and $\{222\}_{pc}$ reflections.

**Figure S4.** Rietveld refined x-ray powder diffraction patterns of KNLNT-$x$ for $x = 0.25$ using **(a)** *Amm2+P4mm* and, **(b)** *P4mm* model. The inset shows magnified view of the fitted $\{200\}_{pc}$ and $\{222\}_{pc}$ reflections.

**Figure S5.** FESEM micrographs of the KNLNT-$x$ for **(a)** $x = 0.05$, **(b)** $x = 0.30$, and **(c)** $x = 0.40$.

**Figure S6.** Histogram images of the distribution of grain size of KNLNT-$x$ for (a) $x = 0$, (b) $x = 0.10$, (c) $x = 0.20$, and (d) $x = 0.30$.

**Figure S7.** Variation of dielectric constant and dielectric loss as a function of temperature at selected frequencies of KNLNT-$x$ for **(a)** $x = 0.05$, **(b)** $x = 0.30$, and **(c)** $x = 0.40$.

**Figure S8.** Magnified plot of $\varepsilon_r$ as a function of temperature around $T_{O-T}$ at selected frequencies of KNLNT-$x$ for (a) $x = 0.00$, (b) $x = 0.05$, (c) $x = 0.10$, and (d) $x = 0.15$.

**Figure S9.** Modified Curie Wiess law plot ($\ln(1/\varepsilon_r - 1/\varepsilon_m)$ versus $\ln(T-T_m)$) of the KNLNT-$x$ for (a) $x = 0$, (b) $x = 0.05$, (c) $x = 0.10$, (d) $x = 0.15$, (e) $x = 0.20$, (f) $x = 0.25$, (g) $x = 0.30$, and (h) $x = 0.40$.

**Figure S10.** Variation of tetragonality with composition ($x$) for KNLNT-$x$ ($0.00 \leq x \leq 0.40$).

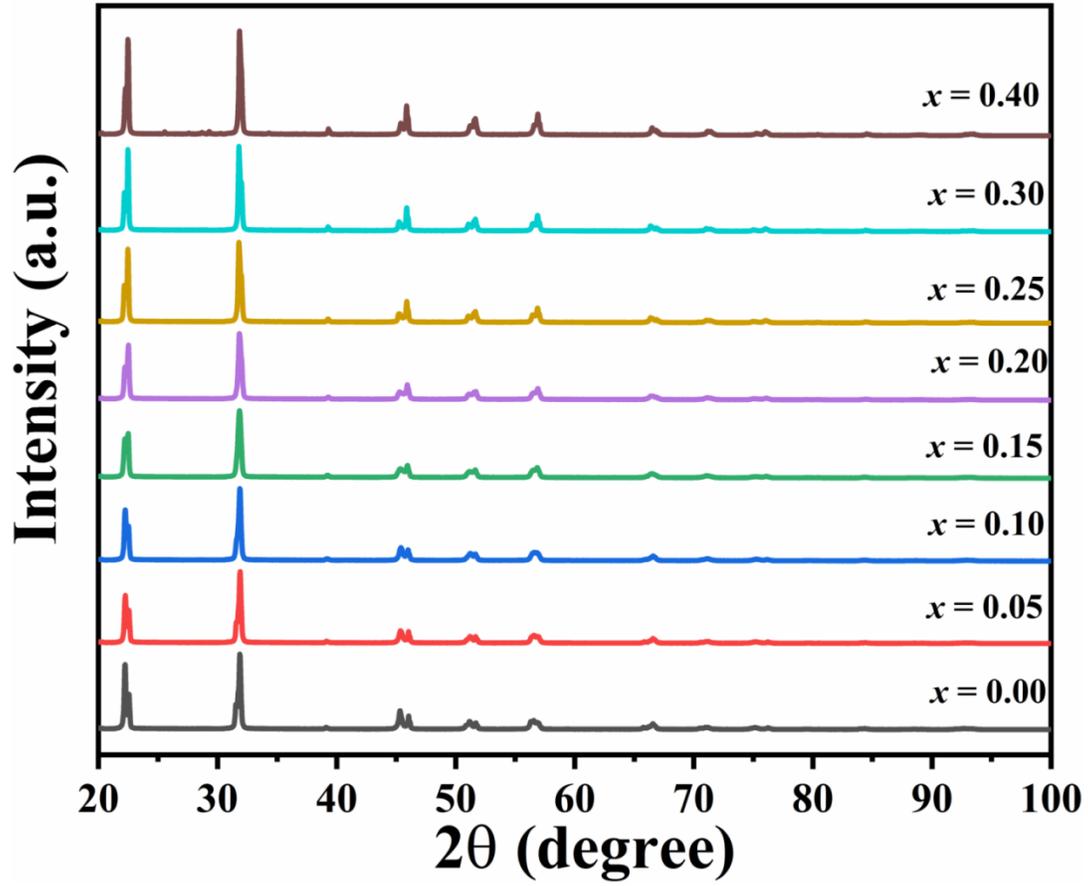

**Figure S1.** Room temperature XRD patterns of KNLNT-*x* with 0 ≤ *x* ≤ 0.4.

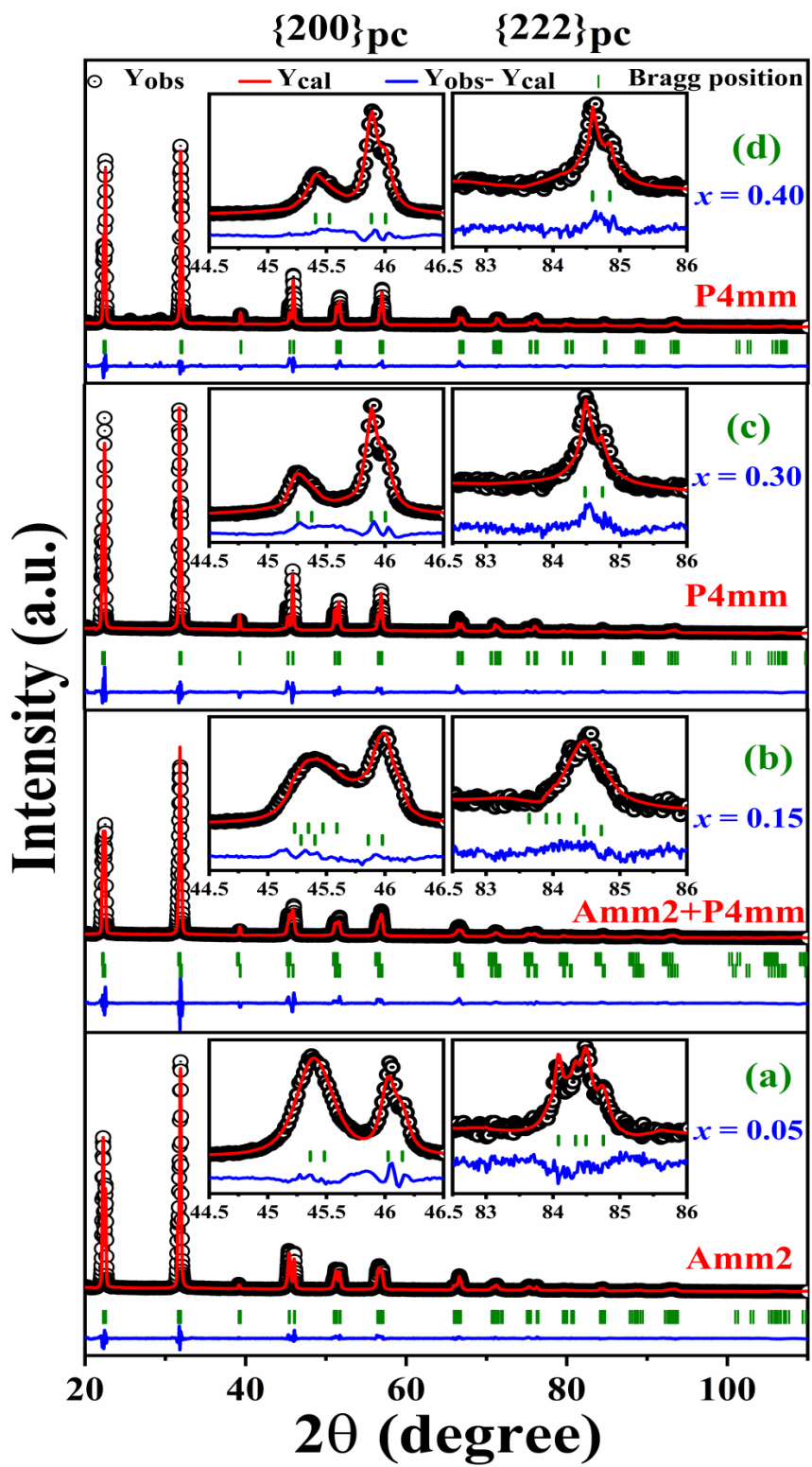

**Figure S2.** Rietveld refined x-ray powder diffraction patterns of KNLNT-$x$ for **(a)** $x = 0.05$ using *Amm2* model, **(b)** $x = 0.15$ using *Amm2+P4mm* model, **(c)** $x = 0.30$ and **(d)** $x = 0.40$ using *P4mm* model. The inset shows magnified view of the fitted $\{200\}_{pc}$ and $\{222\}_{pc}$ reflections.

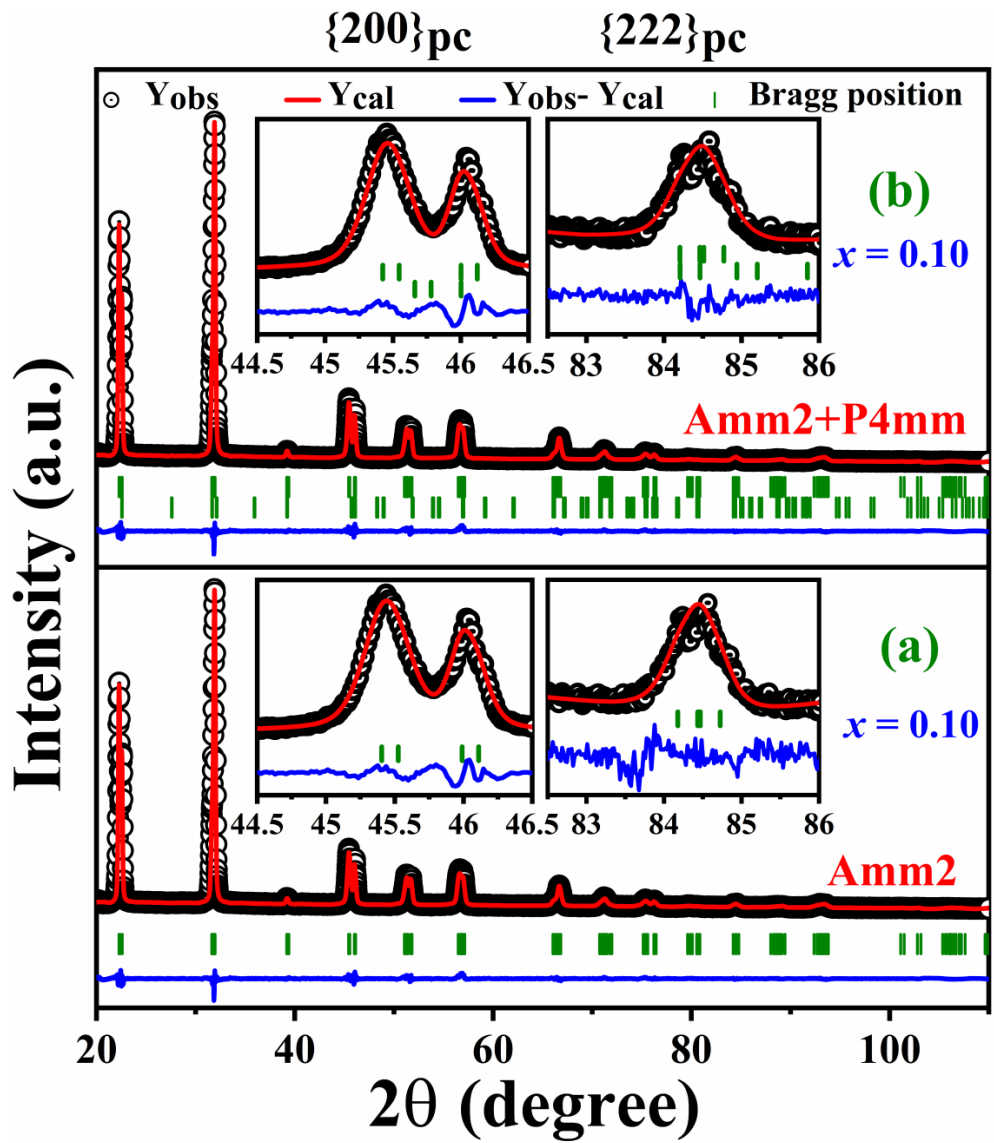

**Figure S3.** Rietveld refined x-ray powder diffraction patterns of KNLNT-$x$ for $x$ = 0.10 using **(a)** *Amm2* and, **(b)** *Amm2+P4mm* model. The inset shows magnified view of the fitted $\{200\}_{pc}$ and $\{222\}_{pc}$ reflections.

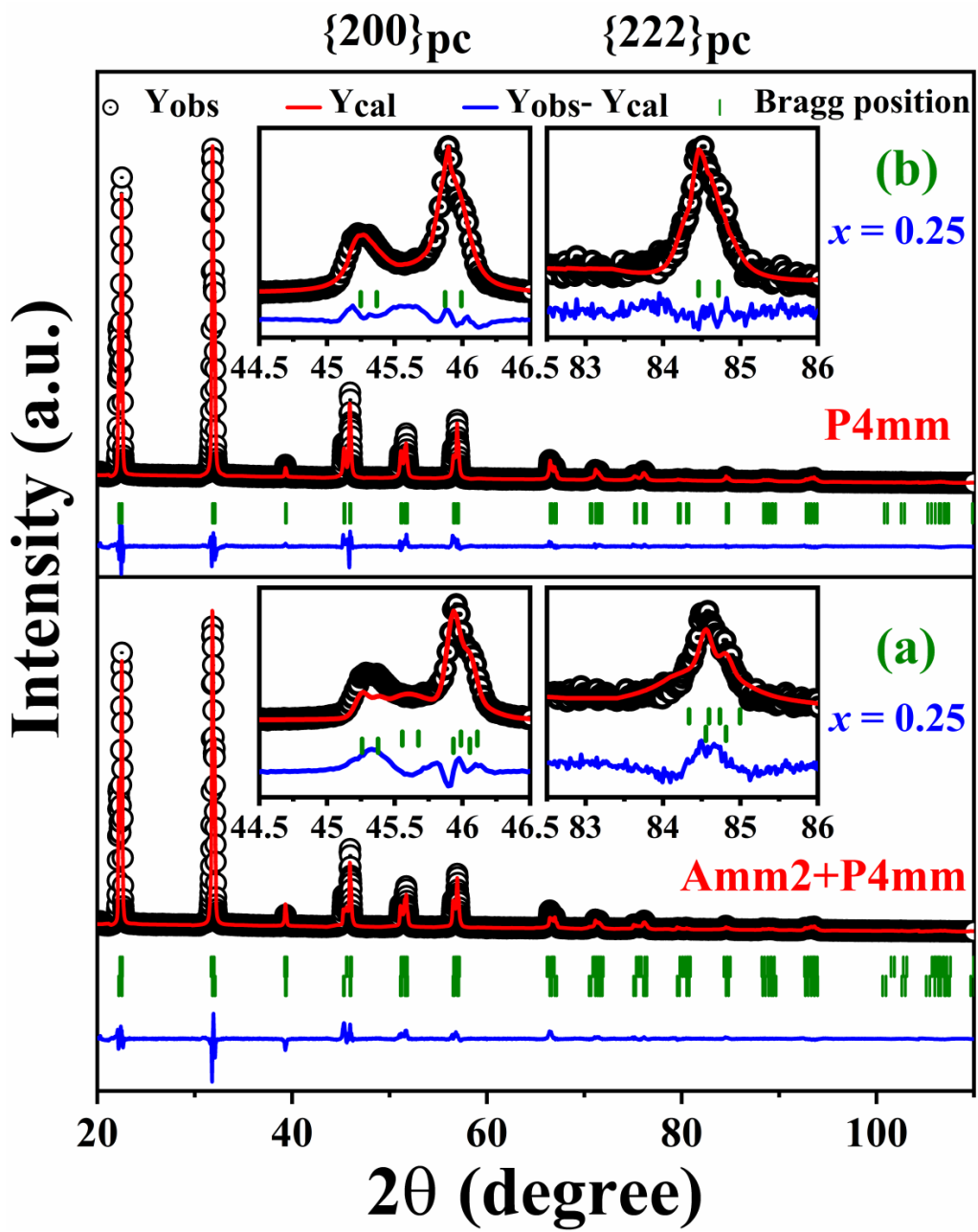

**Figure S4.** Rietveld refined x-ray powder diffraction patterns of KNLNT-$x$ for $x = 0.25$ using **(a)** *Amm2+P4mm* and, **(b)** *P4mm* model. The inset shows magnified view of the fitted $\{200\}_{pc}$ and $\{222\}_{pc}$ reflections.

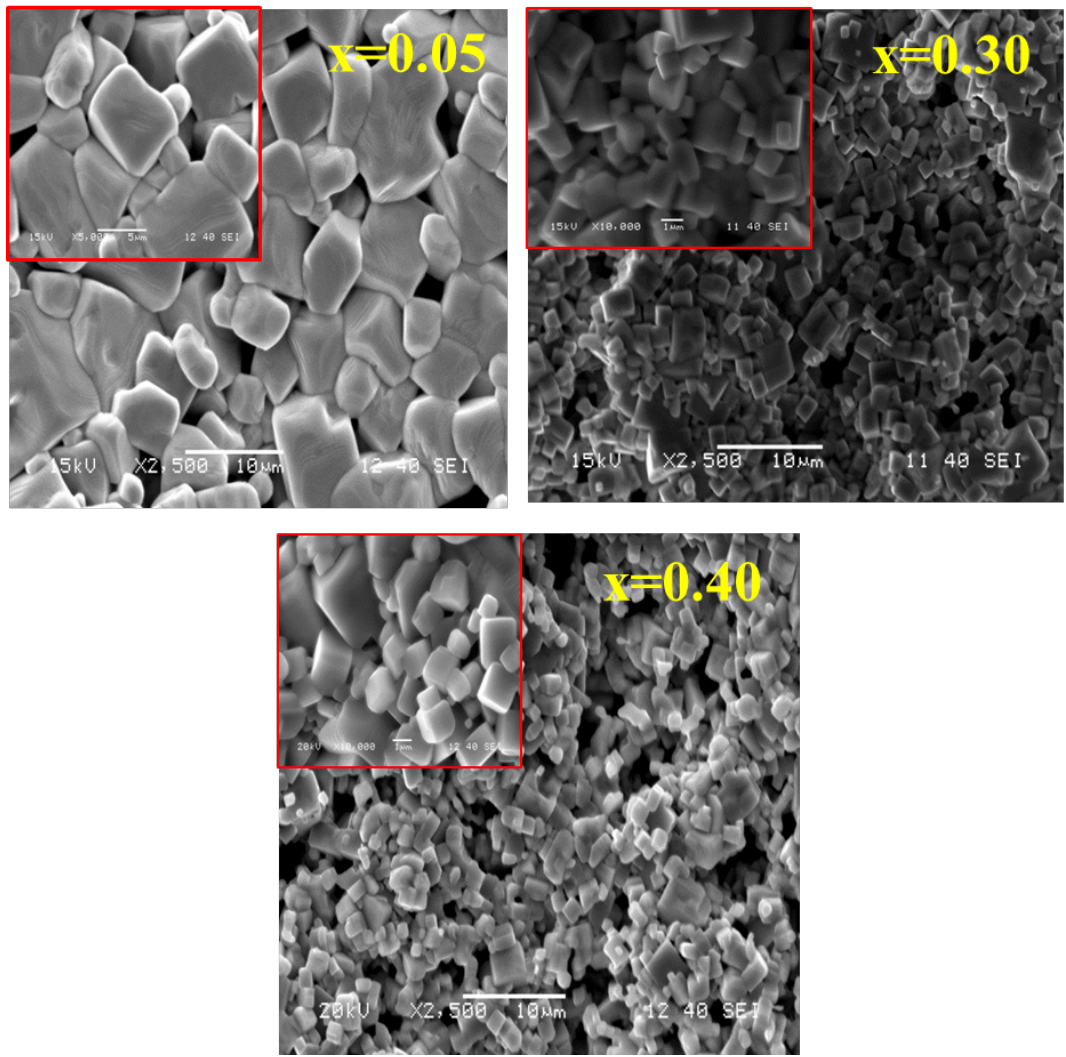

**Figure S5.** FESEM micrographs of the KNLNT-$x$ for **(a)** $x$ = 0.05, **(b)** $x$ = 0.30, and **(c)** $x$ = 0.40.

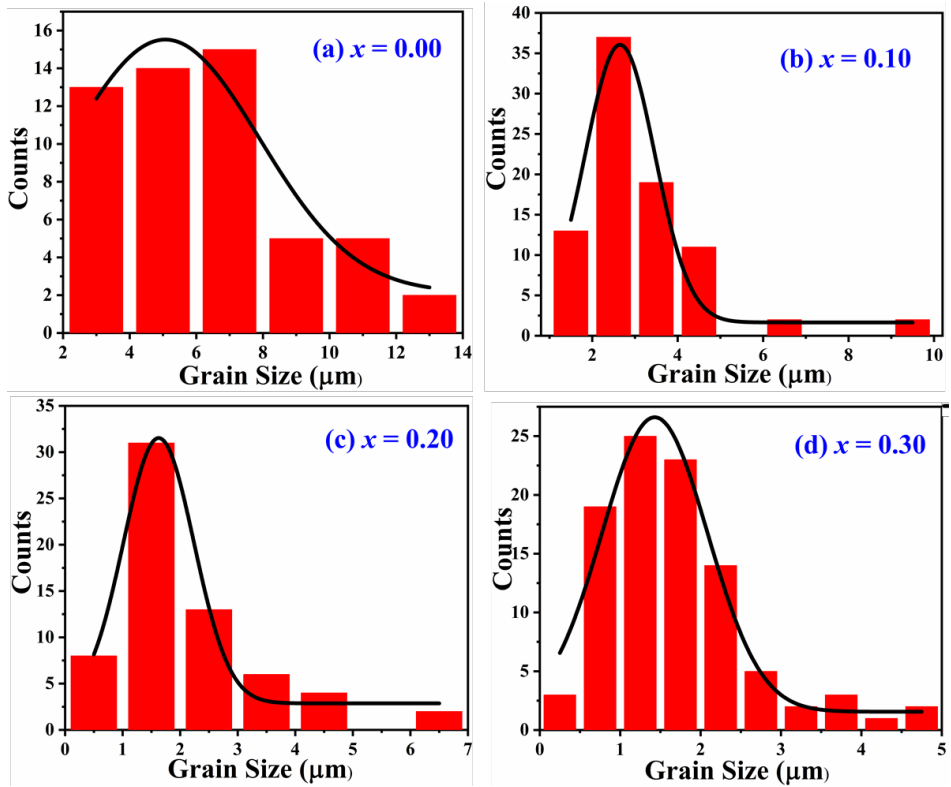

**Figure S6.** Histogram images of the distribution of grain size of KNLNT-$x$ for (a) $x$ = 0, (b) $x$ = 0.10, (c) $x$ = 0.20, and (d) $x$ = 0.30.

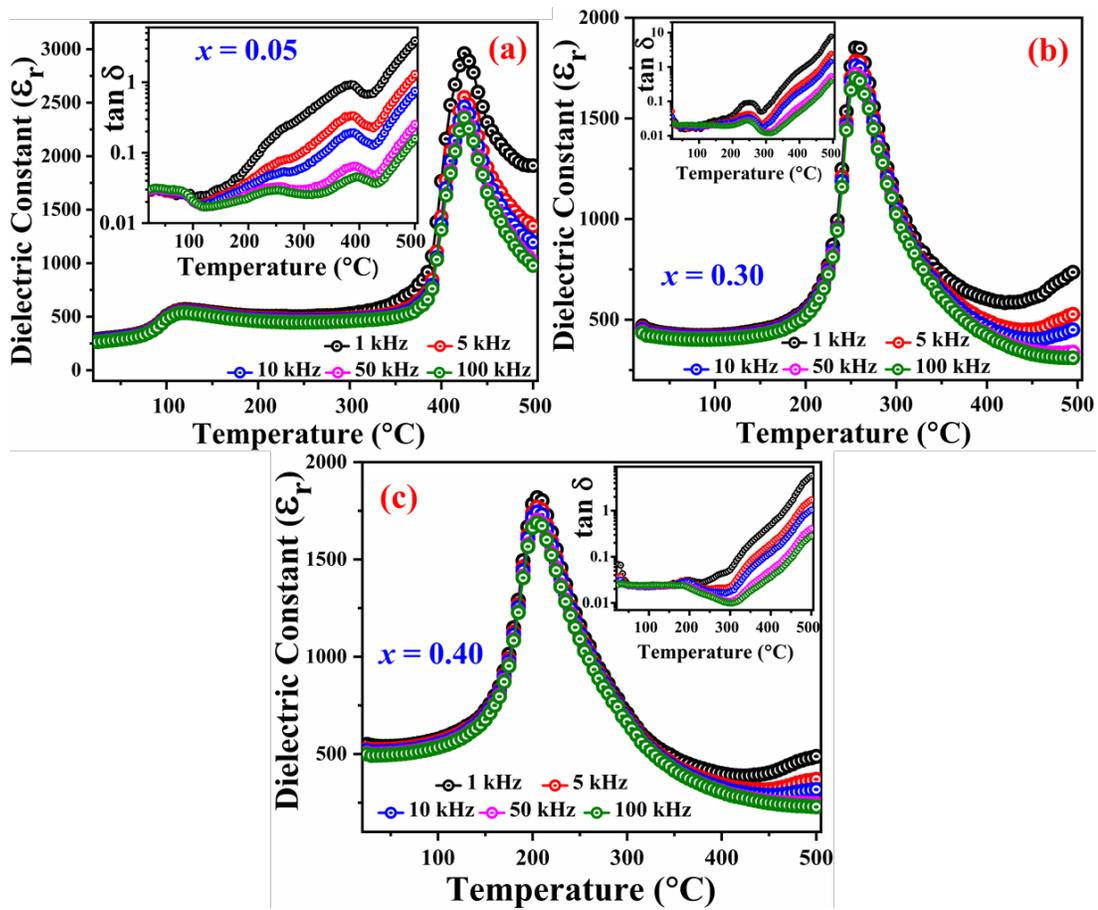

**Figure S7.** Variation of dielectric constant and dielectric loss as a function of temperature at selected frequencies of KNLNT-$x$ for **(a)** $x$ = 0.05, **(b)** $x$ = 0.30, and **(c)** $x$ = 0.40.

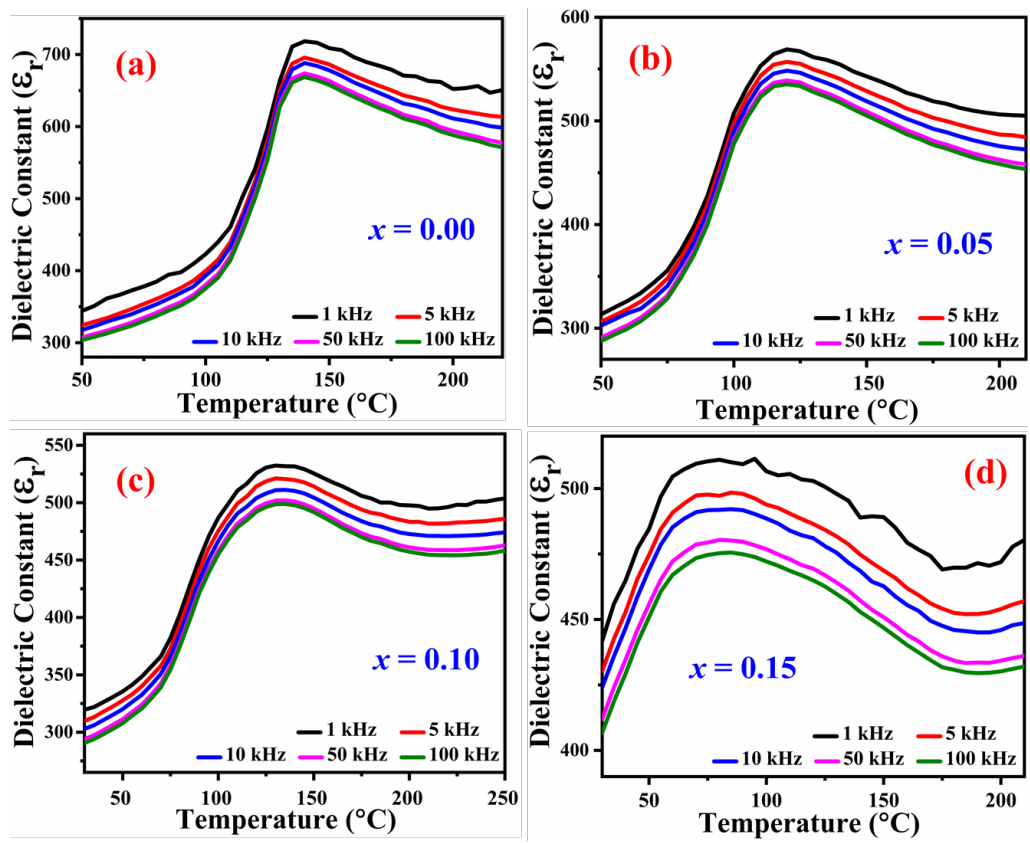

**Figure S8.** Magnified plot of $\varepsilon_r$ as a function of temperature around $T_{O-T}$ at selected frequencies of KNLNT-$x$ for (a) $x = 0.00$, (b) $x = 0.05$, (c) $x = 0.10$, and (d) $x = 0.15$.

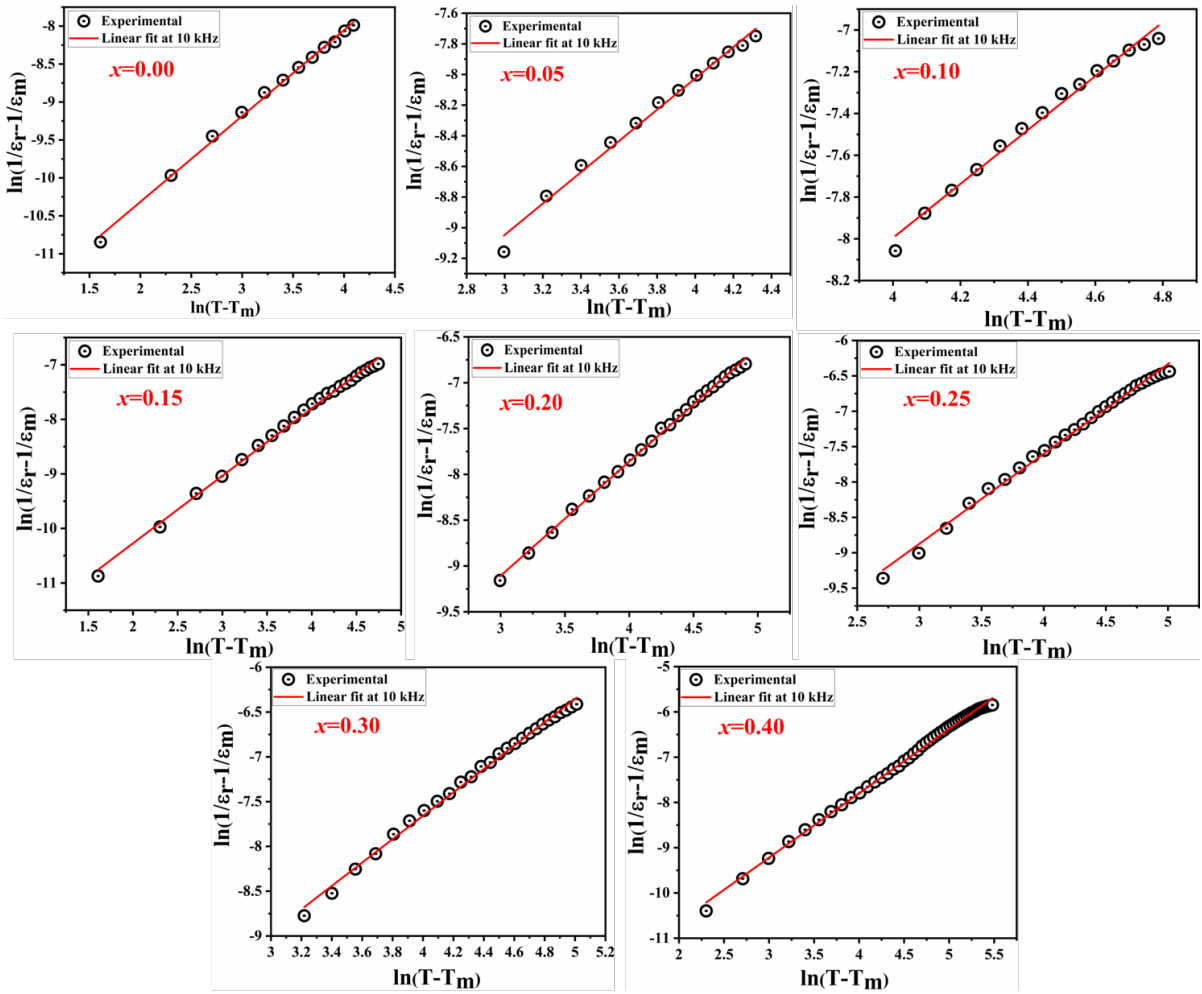

**Figure S9.** Modified Curie Wiess law plot (ln(1/$\varepsilon_r$-1/$\varepsilon_m$) versus ln(T-T$_m$)) of the KNLNT-*x* for (a) *x* = 0, (b) *x* = 0.05, (c) *x* = 0.10, (d) *x* = 0.15, (e) *x* = 0.20, (f) *x* = 0.25, (g) *x* = 0.30, and (h) *x* = 0.40.

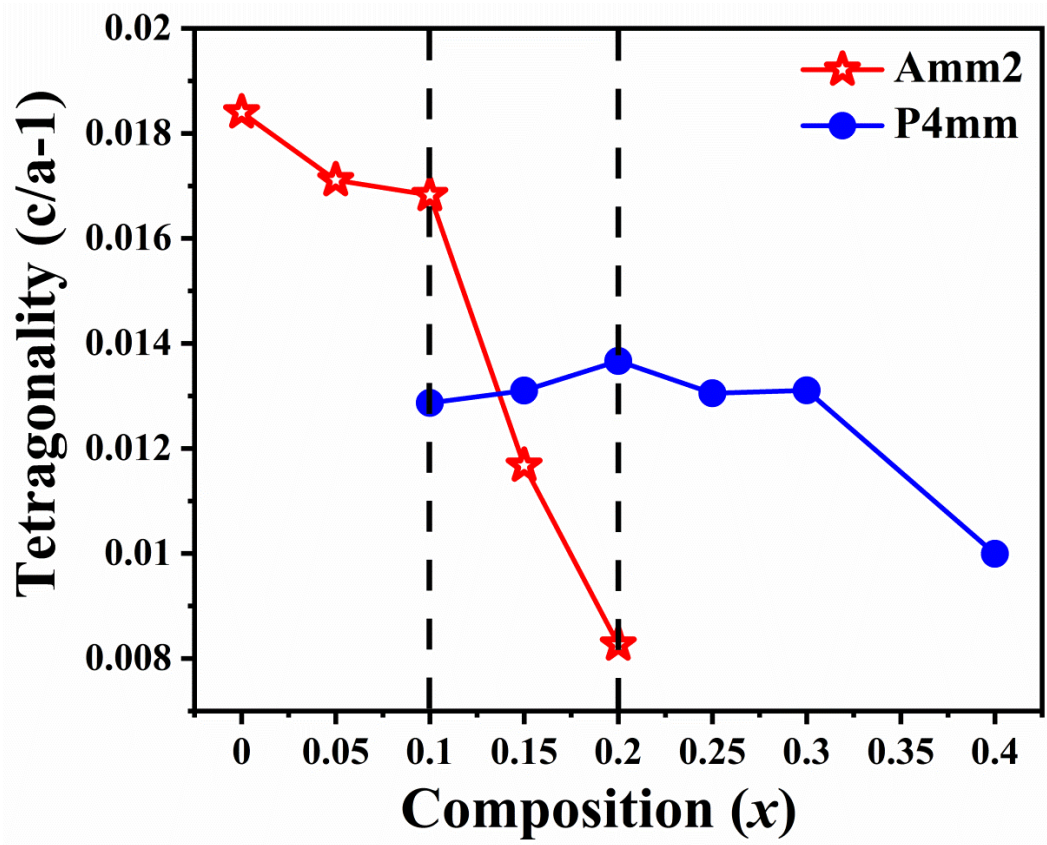

**Figure S10.** Variation of tetragonality with composition ($x$) for KNLNT-$x$ ($0.00 \leq x \leq 0.40$).

## Supplementary Tables:

**Table S1.** A brief review on the crystal structure and physical properties of Li and/or Ta substituted KNN reported by various research groups.

| Sl. No. | Research group | Details of sample composition | Crystal structure with symmetry | Properties | Ref. |
|---|---|---|---|---|---|
| 1 | Mgbemere et al. [8] | $(K_{0.48}Na_{0.48}Li_{0.04})NbO_3$ | Orthorhombic (Amm2) | Li substituted KNN show high dielctric constant, lattice distortion, spontaneous polarization and strain. In Li/Ta co-substituted sample, a wide temperature range of coexistance of orthorhombic and tetragonal phase is observed. | J. Appl. Crys. 44, 1080-1089 (2011) |
|   |   | $(K_{0.5}Na_{0.5})(Nb_{0.9}Ta_{0.1})O_3$ | Rhombohedral (R3c) |  |  |
|   |   | $(K_{0.48}Na_{0.48}Li_{0.04})(Nb_{0.9}Ta_{0.1})O_3$ | Orthorhombic (Amm2) |  |  |
| 2 | Guo et al. [13] | $(1-x)(K_{0.5}Na_{0.5})NbO_3-xLiNbO_3$ | Orthorhombic for $x \leq 0.05$ | High Curie temperature (>450 °C) has been reported. The ceramics show enhanced piezoelectric and electromechanical responses ($d_{33}$=200–235 pC/N, $k_p$=38%–44%, and $k_t$=44%–48%) in the MPB compositions. | Appl. Phys. Lett. 85, 4121-4123 (2004) |
|   |   |   | Orthorhombic+Tetragonal for $0.05<x<0.07$ |  |  |
|   |   |   | Tetragonal for $x \geq 0.07$ |  |  |
| 3 | Wang et al. [14] | $(1–x)(Na_{0.535}K_{0.48})NbO_3–xLiNbO_3$ ($x$ = 0.058–0.085) | Orthorhombic+Tetragonal ($x$ = 0.073–0.083) | Highest $d_{33}$ of 324 pC/N ($x$ = 0.080) was obtained. | Adv. Funct. Mater. 20, 1924-1929 (2010) |
| 4 | Hollenstein et al. [15] | $(K_{0.5-x/2},Na_{0.5-x/2},Li_x)NbO_3$ |  | $T_C$ shifted to higher temperatures and $T_{O-T}$ shifted towards RT with increase in Li content and for $x$ = 0.07, $T_{O-T}$ shifted below RT indicating lithium addition stabilizes the tetragonal phase. | Appl. Phys. Lett. 87, 182905 (2005). |
|   |   | $(K_{0.5-x/2},Na_{0.5-x/2},Li_x)(Nb_{1-y}Ta_y)O_3$ |  |  |  |

| # | Author | Composition | Phase | Key findings | Reference |
|---|---|---|---|---|---|
| 5 | Dai et al. [17] | $(1-x)(K_{0.5}Na_{0.5}NbO_3) - xLiTaO_3$ | Orthorhombic+ Tetragonal ($x = 0.05$) | Li-substituted ceramics shows $k_t$ of 53% and converse piezoelectric coefficient $d_{33}$ around 200 pm/V while a $kt$ of 52% and $d_{33}$ over 300 pm/V is observed for Li- and Ta-modified samples. The phase coexistence (O and T) is due to the shifting of $T_{O-T}$ aound RT and not due to the MPB. | Appl. Phys. Lett. 90, 262903 (2007) |
| 6 | Wang et al. [18] | $(1-x)(K_{0.5}Na_{0.5}NbO_3) - xLiNbO_3$ ($0 \leq x$ (mol%) $\leq 8$) | Orthorhombic ($x = 0$) Tetragonal ($x = 0.08$) | Enhanced piezoelctric constant $d_{33}$ = 215 pC/N was observed for $x$ = 6 mol% | Appl. Phys. Lett. 91, 262902 (2007) |
| 7 | Wu et al. [20] | $(K_xNa_{0.96-x}Li_{0.04})(Nb_{0.91}Ta_{0.05}Sb_{0.04})O_3$ ($x$ = 0.32–0.56) | Orthorhombic at $x \leq 0.37$ Tetragonal at $x > 0.39$ Orthorhombic + Tetragonal $0.37 \leq x \geq 0.39$ | The ceramics with x=0.38 exhibit enhanced electrical properties ($d_{33}$ ~ 306 pC/N; $k_p$ ~ 48%; $k_t$ ~ 49%; $T_c$ ~ 337 °C; $\varepsilon_r$ ~ 1327; tan δ ~ 2.5%; $P_r$ ~ 34.9 μC/cm²; Ec ~ 11.3 kV/cm. | Appl. Phys. Lett. 91, 252907 (2007). |
| 8 | Sung et al. [21] | $(K_{0.53}Na_{0.47-x}Li_x)$ and $(K_{0.53}Na_{0.47})(Nb_{1-y}Ta_y)O_3$ $0 \leq x \leq 0.09$ and $0 \leq y \leq 0.06$ | Orthorhombic+ Tetragonal ($x$ = 0.06 and $y$ = 0.45) | The ceramics with $x$=0.06 and y=0.45 shows enhanced electrical properties ($d_{33}$=251 pC/N, $k_p$ =0.49, $\varepsilon_3^T/\varepsilon_0$=1260, tan δ = 0.03 and $T_C$ =376 °C) owing to the formation of a MPB. | Appl. Phys. Lett. 105, 142903 (2014). |
| 9 | Wang et al. [22] | $(Li_{0.04}Na_{0.48}K_{0.48})(Nb_{0.80}Ta_{0.20})O_3$ | Orthorhombic at room temperature | $T_C$ decreases and $T_{O-T}$ increases with increase of annealing time. The room temperature dielectric, ferroelectric, and piezoelectric properties are quite sensitive to the growth of the abnormal grains. | J. Am. Ceram. Soc. 91, 1962-1970 (2008). |

| # | Author | Composition | Phase | Key findings | Reference |
|---|---|---|---|---|---|
| 10 | Chang et al. [29] | $(K_xNa_{0.96-x}Li_{0.04})(Nb_{0.85}Ta_{0.15})O_3$ | Orthorhombic ($x = 0.36$) <br><br> Orthorhombic + Tetragonal ($0.42 \leq x \leq 0.46$) <br><br> Tetragonal ($x = 0.52$) | The ceramic with $x = 0.44$ shows optimum electrical properties ($d_{33} = 291$ pC/N, $K_p = 0.54$, $\varepsilon_r = 1167$, $T_{O-T} = 35$ °C, $T_C = 351$ °C, $P_r = 27.65$ μC/cm$^2$, $E_C = 8.63$ kV/cm). | J. Appl. Phys. 105, 054101 (2009) |
| 11 | Klein et al. [30] | $(1-x)(Na_{0.5}K_{0.5})NbO_3-xLiNbO_3$ ($x = 0$ to $0.1$) | Orthorhombic ($0 \leq x \leq 0.07$) <br> Tetragonal ($x > 0.07$) | Both the vertical MPB and the triple point (T+O+M) are observed around ($x = 0.07$) regions where enhanced piezoelectric and dielectric properties are obtained. | J. Appl. Phys. 102, 014112 (2007). |
| 12 | Ge et al. [31] | <001> textured $(K_{0.5}Na_{0.5})_{0.98}Li_{0.02}NbO_3$ | Orthorhombic at RT | Increasing E along the texture direction resulted in a notable increase in the volume fraction of the T phase at the expense of the O phase which enhances piezoelectric properties. | Phys. Rev. B 83, 224110 (2011) |
| 13 | Zhang et al. [32] | $(K_{0.5}Na_{0.5})_{0.94}Li_{0.06}NbO_3$ (KNLN) and $(K_{0.5}Na_{0.5})Li_{0.04}(Nb_{0.85}Ta_{0.15})O_3$ (KNLNT) | Tetragonal at room temperature (RT). | The orthorhombic-tetragonal phase transition for both the compositions is close to RT. <br><br> They shows high piezoelctrcity (KNLN=225 pC/N, KNLN=258 pC/N) at RT. | J. Appl. Phys. 116, 104106 (2014) |
| 14 | Yao et al. [33] | $Li_{0.03}(K_{0.48}Na_{0.52})(Nb_{0.8}Ta_{0.2})O_3$ | Orthorhombic | The ceramic shows elastically harder than conventional PZT Possess higher piezoelctric stiffness conastant ($h_{33} = 68.88*10^8$ V/m) and large electro mechanical coupling factor ($K_{33} = 57\%$) | J. Appl. Phys. 113, 174105 (2013) |
| 15 | Zhang et al. [34] | $(K_{0.55}Na_{0.45})_{0.965}Li_{0.035}Nb_{0.80}Ta_{0.20}O_3$ | Orthorhombic + Tetragonal structure | Possess highest $d_{33} = 262$ pC/N, $K_p = 0.53$, $K_{33} = 0.63$, $K_{31} = 0.31$, $\varepsilon = 1290$ at RT. | Appl. Phys. Lett. 95, 022909 (2009) |
| 16 | Kong et al. [35] | $(1-x)(Na_{0.5}K_{0.5})NbO_3-xLiNbO_3$ | monoclinic to tetragonal with MPB at $x = 0.06$ (as | The monoclinic to tetragonal phase transition belongs to order-disorder type. | Inorg. Chem. 61, |

| | | ($x$ = 0.00–0.08) | per neutron Bragg diffraction data). | | 4335-4349 (2022) |
|---|---|---|---|---|---|
| | | | Monoclinic for $0 \leq x \leq 0.08$ as per neutron pair distribution analysis | | |
| 17 | Ge et al. [36] | $(K_{0.5}Na_{0.5})NbO_3$-5%$LiNbO_3$ | Monoclinic at RT | Transition of Monoclinic ($M_C$) phase to a tetragonal (P4mm) on heating between 340 K and 360 K indicating a means to achieve high piezoelectricity | J. Appl. Phys. 111, 103503 (2012). |
| 18 | Wang et al. [37] | $(Li_{0.04}Na_{0.48}K_{0.48})(Nb_{0.80}Ta_{0.20})O_3$ | | In direct mixing, inhomogeneous distribution of A- and B-site cations is found. In precursor method, improved compositional homogeneity and appreciably enhanced dielectric, ferroelectric, and piezoelectric properties are obtained. | J. Am. Ceram. Soc. 90, 3485-3489 (2007) |

**Table S2**. Refined structural parameters obtained from Rietveld refinement of KNLNT-$x$ ceramics for $0 \leq x \leq 0.4$.

| Composition $x$ | Space Group | $\chi^2$ | a (Å) | b (Å) | c (Å) | Volume (Å$^3$) | Phase Fraction (%) |
|---|---|---|---|---|---|---|---|
| 0.00 | Amm2 | 3.45 | 3.9396 | 5.6385 | 5.6719 | 125.99 | 100 (O) |
| 0.05 | Amm2 | 2.04 | 3.9406 | 5.6331 | 5.6671 | 125.80 | 100 (O) |
| 0.10 | Amm2+P4mm | 4.28 | 3.9420 | 5.6290 | 5.6604 | 125.60 | 83.43 (O) |
|  |  |  | 3.9631 | 3.9631 | 4.0141 | 63.05 | 16.57(T) |
| 0.15 |  | 10.7 | 3.9615 | 5.6699 | 5.6510 | 126.93 | 74.76(O) |
|  | Amm2+P4mm |  | 3.9691 | 3.9691 | 3.9925 | 62.89 | 25.24(T) |
| 0.20 |  | 2.71 | 3.9468 | 5.6594 | 5.6269 | 125.69 | 44.34(O) |
|  | Amm2+P4mm |  | 3.9520 | 3.9520 | 4.0060 | 62.57 | 55.66(T) |
| 0.25 | P4mm | 6.35 | 3.9532 | 3.9532 | 4.0048 | 62.58 | 100 (T) |
| 0.30 | P4mm | 8.98 | 3.9523 | 3.9523 | 4.0041 | 62.55 | 100 (T) |
| 0.40 | P4mm | 10.3 | 3.9522 | 3.9522 | 3.9915 | 62.34 | 100 (T) |

**Table S3**. Variation of grain size, $\gamma$, $\varepsilon_r$, $2P_r$, $E_c$, and $d_{33}$ measured at RT with composition of KNLNT-$x$ ($0 \leq x \leq 0.4$).

| Composition ($x$) | Grain Size (μm) | $\gamma$ | $\varepsilon_r$ at RT | $2P_r$ (μC/cm$^2$) | $E_c$ (kV/cm) | $d_{33}$ (pC/N) |
|---|---|---|---|---|---|---|
| 0.00 | 6.43 | 1.21 | 294 | 69.11 | 12.18 | 70 |
| 0.05 | 6.17 | 1.25 | 278 | 39.50 | 10.67 | 75 |
| 0.10 | 3.13 | 1.32 | 309 | 30.59 | 10.04 | 108 |
| 0.15 | 2.49 | 1.37 | 427 | 35.92 | 9.81 | 118 |
| 0.20 | 2.02 | 1.41 | 556 | 40.62 | 9.3 | 159 |
| 0.25 | 1.82 | 1.42 | 480 | 33.85 | 9.52 | 93 |
| 0.30 | 1.70 | 1.45 | 451 | 10.28 | 9.95 | 89 |
| 0.40 | 1.89 | 1.52 | 518 | 21.21 | 11.06 | 51 |